\newif\ifshortver
\shortvertrue
\shortverfalse

\newif\iflongver
\ifshortver
\documentclass[conference,letterpaper]{IEEEtran}
\longverfalse
\newcommand{\myskip}{\smallskip}
\newcommand{\mysmallskip}{}
\else
\documentclass[english, onecolumn]{IEEEtran}
\longvertrue

\newcommand{\myskip}{\medskip}
\newcommand{\mysmallskip}{\medskip}
\fi

\ifshortver
\fi

\usepackage[utf8]{inputenc} 
\usepackage[T1]{fontenc}
\usepackage{url}
\usepackage{ifthen}
\usepackage{cite}
\usepackage[cmex10]{amsmath}

\usepackage{algorithm}
\usepackage[noend]{algpseudocode}
\algrenewcommand\algorithmicindent{5mm}

\usepackage{amssymb}
\usepackage{amsfonts}
\usepackage{cite}
\usepackage{tikz}
\usepackage{bbm}
\usetikzlibrary{positioning,arrows,shapes,chains,fit,scopes}
\usepackage{pgfplots}
\usepackage{pgfplotstable}	
\usepackage{booktabs}
\usepackage{colortbl}
\usepackage{color}
\usepackage{blkarray, bigstrut}
\usepackage{amsmath}

\IEEEoverridecommandlockouts

\newtheorem{theorem}{Theorem}
\newtheorem{lemma}[theorem]{Lemma}
\newtheorem{corollary}[theorem]{Corollary}

\newtheorem{definition}{Definition}

\title{One-Shot Coding over General Noisy Networks}

\begin{document}

\author{
Yanxiao Liu, \textit{Graduate Student Member, IEEE}, and Cheuk Ting Li, \textit{Member, IEEE}
\thanks{
This work was partially supported by two grants from the Research Grants Council of the Hong Kong Special Administrative Region, China [Project No.s: CUHK 24205621 (ECS), CUHK 14209823 (GRF)].

This paper was presented in part at the 2024 IEEE International Symposium on Information Theory (ISIT) [DOI:10.1109/ISIT57864.2024.10619467]. 

Yanxiao Liu and Cheuk Ting Li are with the Department of Information Engineering, The Chinese University of Hong
Kong, Hong Kong SAR of China. Email:  yanxiaoliu@link.cuhk.edu.hk,  ctli@ie.cuhk.edu.hk. 
}
}

\maketitle

\begin{abstract}
We present a unified one-shot coding framework designed for the communication and compression of messages among multiple nodes across a general acyclic noisy network. Our setting can be seen as a one-shot version of the acyclic discrete memoryless network studied by Lee and Chung, and noisy network coding studied by Lim, Kim, El Gamal and Chung. We design a proof technique, called the exponential process refinement lemma, that is rooted in the Poisson matching lemma by Li and Anantharam, and can significantly simplify the analyses of one-shot coding over multi-hop networks. Our one-shot coding theorem not only recovers a wide range of existing asymptotic results, but also yields novel one-shot achievability results in different multi-hop network information theory problems, such as compress-and-forward and partial-decode-and-forward bounds for a one-shot (primitive) relay channel, and a bound for one-shot cascade multiterminal source coding. In a broader context, our framework provides a unified one-shot bound applicable to any combination of source coding, channel coding and coding for computing problems. 
\end{abstract}

\iflongver
\section{Introduction}
\label{sec::intro}
In information theory, to find optimal and reliable information transmission rates in networks, we often employ the asymptotic equipartition property and the law of large numbers to study the asymptotic behavior of channels and networks in the large blocklength limit, as in conventional asymptotic typicality-based proofs such as~\cite{el2011network}. 
However, due to the fact that packet lengths are not infinite in real applications (they can even be very short in ultrareliable low-latency communications~\cite{durisi2016toward}), \emph{finite blocklength information theory} has been widely studied in the past decade. 
The goal is to provide nonasymptotic results of channel capacities in the scenario where the number of channel uses is limited. 
See~\cite{kostina2013lossy,  wang2011dispersion, tan2013dispersions, polyanskiy2010channel, liu2024oblivious_isit} and references therein.

In particular, in a \emph{one-shot coding} setting \cite{feinstein1954new, shannon1957certain, hayashi2009information, polyanskiy2010channel, verdu2012non, yassaee2013technique, liu2015one, song2016likelihood, watanabe2015nonasymptotic, yassaee2013non, li2021unified}, we assume the blocklength is $1$, i.e., each source and channel can be only used \emph{once}. 
One-shot settings are general, in the sense that the sources and channels can be arbitrary, and does not have to be memoryless or ergodic.
The goal is to derive one-shot achievability results that can imply existing (first-order and second-order) asymptotic results when applied to memoryless sources and channels, or applied to systems with memory that behave ergodically~\cite{verdu1994general}. 
For example, for point-to-point channel coding, the achievability of the channel capacity is implied by the one-shot bounds by Feinstein~\cite{feinstein1954new}, Shannon~\cite{shannon1957certain}, and the bounds in \cite{hayashi2009information,polyanskiy2010channel}.

We briefly review existing one-shot results for multi-user coding settings. 
In~\cite{verdu2012non}, 
one-shot versions of the covering and packing lemmas have been proposed and applied to various problems in multiuser information theory, for example, multiple access channels and broadcast channels. 
In~\cite{yassaee2013technique}, a proof technique based on stochastic likelihood encoders and decoders has been used to derive various one-shot achievability results in several multi-user settings, including broadcast channels, multiterminal source coding and multiple description coding. 
A one-shot mutual covering lemma has been proposed in~\cite{liu2015one} for broadcast channels, which recovers Marton's inner bound. 
In~\cite{song2016likelihood}, the multiterminal source coding inner bound has been examined by a likelihood encoder. 
A finite-blocklength version of the random binning technique has been used in~\cite{yassaee2013non} to derive second order regions for broadcast channels. 
Recently, in~\cite{li2021unified}, a technique called the \emph{Poisson matching lemma} has been introduced to prove various one-shot achievability results for a range of coding settings, and the achievable one-shot bounds improve upon the best known one-shot bounds in several settings with shorter proofs. This technique has been applied to the information hiding problem~\cite{liu2024hiding}, unequal message protection~\cite{khisti2024unequal}, hypothesis testing~\cite{guo2024hypothesis} and secret key generation~\cite{hentila2024communication}.
The Poisson matching lemma is based on the \emph{Poisson functional representation}~\cite{li2018strong}. 
Together with other related techniques, the Poisson functional representation~\cite{li2018strong} has been applied to various fields recently, e.g., neural estimation~\cite{lei2022neural}, minimax learning~\cite{li2020minimax} and differential privacy~\cite{liu2024universal}. 
 
In this paper, which is the complete version of~\cite{liu2024one_isit},\footnote{The conference paper~\cite{liu2024one_isit} includes the description (but not the complete proof) of our one-shot achievable region and the corresponding coding schemes. 
Compared to~\cite{liu2024one_isit}, this complete version includes the complete analysis and proof of our one-shot achievability results, as well as the proof of the lemma needed and the additional explanations. 
Moreover, this complete version also discusses a novel one-shot cascade multiterminal source coding with computing problem, which has not been discussed in~\cite{liu2024one_isit}. 
Additional figures have also been provided for the purpose of illustration.}
we study a general class of networks, which we call \emph{acyclic discrete networks}, where there are $N$ nodes connected by noisy channels in an acyclic manner. Each node can play the role of an encoder or a decoder (or both) in source coding or channel coding settings. This is a one-shot version of the asymptotic acyclic discrete memoryless network studied by Lee and Chung~\cite{lee2018unified}, and includes a wide range of settings as special cases, such as source and channel coding, primitive relay channel~\cite{el2011network, kim2007coding, mondelli2019new, el2021achievable, el2022strengthened}, Gelfand-Pinsker~\cite{gelfand1980coding,Heegard1980}, relay-with-unlimited-look-ahead~\cite{el2005relay, el2007relay}, Wyner-Ziv~\cite{wyner1976rate, wyner1978rate}, coding for computing~\cite{yamamoto1982wyner},  multiple access channels~\cite{ahlswede1971multi, liao1972multiple, ahlswede1974capacity}, broadcast channels~\cite{marton1979coding} and cascade multiterminal source coding~\cite{cuff2009cascade}. In a broader context, our one-shot achievability results are general enough to be applicable to any
combination of source coding, channel coding and coding for computing problems.

In order to alleviate the difficulty of keeping track of a large number of auxiliary random variables in a general $N$-node network, we propose a tool called the \emph{exponential process refinement lemma} based on the Poisson matching lemma~\cite{li2021unified},\footnote{We only present the discrete case in this paper for the sake of simplicity. Hence, instead of Poisson processes, we may use an i.i.d. exponential processes 
instead~\cite{li2018strong}. While we expect the results to be extended to the continuous case, this is left for future studies.} 
which simplifies the analyses of the evolution of the posterior distribution of the sources, messages and/or auxiliary random variables at the decoder.
We utilize the lemma to prove a one-shot achievability result for general acyclic discrete networks, which recovers existing one-shot results in a range of settings in~\cite{li2021unified, verdu2012non, yassaee2013technique, watanabe2015nonasymptotic}, and also give novel one-shot results for various multi-hop settings, namely primitive relay channels~\cite{el2011network, kim2007coding, mondelli2019new, el2021achievable, el2022strengthened}, relay-with-unlimited-look-ahead~\cite{el2005relay, el2007relay}, and cascade multiterminal source coding~\cite{cuff2009cascade}.

One limitation of the acyclic discrete network is that the nodes must be arranged in an acyclic manner, and each node only transmits once. Therefore, it is unable to model cyclic networks such as the two-way communication channel~\cite{shannon1961two} and noisy network coding~\cite{lim2011noisy}, and settings where the transmission occurs in an iterative manner such as the relay channel~\cite{el2011network, van1971three, cover1979capacity} (where the relay chooses each output symbol based on its past received symbols). We argue that this is an inherent limitation of one-shot settings, and the acyclic discrete network already provides the ``best one-shot approximation'' of noisy network coding~\cite{lim2011noisy} and relay channels~\cite{el2011network, van1971three, cover1979capacity} . 
Note that the primitive relay channel~\cite{el2011network, kim2007coding, mondelli2019new, el2021achievable, el2022strengthened} and the relay-with-unlimited-look-ahead~\cite{el2005relay, el2007relay} can be modeled as acyclic discrete networks, since the relay in these settings does not need to transmit in an iterative manner.

The paper is organized as follows. 
We present our proof technique, called the exponential process refinement lemma, in Section~\ref{sec:expon}. 
We describe our general acyclic discrete network in Section~\ref{sec::net_model}, and prove our main theorem in Section~\ref{sec::main}. 
In Section~\ref{sec::relay}, we use a one-shot relay channel and related settings to elaborate our coding scheme in detail. 
We then discuss a novel one-shot cascade multiterminal source coding problem in Section~\ref{sec::cascade}. 
We also show our coding scheme provides one-shot bounds on various network information theory settings in Section~\ref{sec::NIT}. 
The proof of Theorem~\ref{thm::network_achievability} and Theorem~\ref{thm:det}\label{pf:thm} can be found in the appendix. 
 
\fi 

\subsection*{Notations}
We assume all random variables are from finite alphabets and logarithm and entropy are to the base $2$ unless otherwise stated. 
Logarithm to the base $e$ is denoted as $\ln(x)$.
Write $[i..j] := \{i,i+1,\ldots,j \}$, $[j] := [1..j]$.
For a set $\mathcal{S} \subseteq [k]$ and random sequence $U_1,\ldots,U_k$, write $U^k := (U_1,\ldots,U_k)$, $U_\mathcal{S}:= (U_j)_{j\in \mathcal{S}}$. 
For a statement $S$, $\mathbf{1}\{S\}$ is its indicator, i.e., $\mathbf{1}\{S\}$ is $1$ if $S$ holds and $0$ otherwise.
$\iota_{X;Y}(x;y):=\log (P_{X,Y}(x,y)$$/(P_{X}(x)P_{Y}(y)))$ denotes the information density, and
$\iota_{X;Y|Z}(x;y|z):=\log (P_{X,Y|Z}(x,y|z)$$/(P_{X|Z}(x|z)P_{Y|Z}(y|z)))$ denotes the conditional information density. 
We use $\iota(X;Y) $ instead of $\iota_{X;Y}(X;Y)$ when the random variables are clear from the context. 
 $\delta_{a}$ denotes the degenerate distribution $\mathbf{P}\{X=a\}=1$.

\section{Preliminaries}
\label{sec::pre}

In this section, we  review the Poisson functional representation~\cite{li2018strong} and the Poisson matching lemma~\cite{li2021unified}.

\medskip

\begin{definition}[Poisson functional representation]
Consider a finite set $\mathcal{U}$. Let $\mathbf{U}:=(Z_{u})_{u\in\mathcal{U}}$, where the $Z_u$'s
are i.i.d. $\mathrm{Exp}(1)$ random variables.
Given a distribution $P$ over $\mathcal{U}$, the random variable
\begin{equation}
\mathbf{U}_{P} :=\mathrm{argmin}_{u}\frac{Z_{u}}{P(u)} \label{eq:pfr}
\end{equation}
is called the Poisson functional representation \cite{li2018strong}.\footnote{When the space $\mathcal{U}$ is continuous, 
a Poisson process is used in~\cite{li2018strong, li2021unified}, instead of i.i.d. $\mathrm{Exp}(1)$ random variables.
}
\end{definition}

\medskip

The Poisson functional representation ensures that $\mathbf{U}_{P} \sim P$, i.e., it outputs a sample that \emph{exactly} follows the desired distribution. 
This exactness arises from a property of exponential random variables: for positive constants $r_1,r_2$, if $Z_1,Z_2 \sim \mathrm{Exp}(1)$ are i.i.d., then $\mathbf{P}(Z_1/r_1 < Z_2/r_2) = r_1 / (r_1 + r_2)$. 
By generalizing this to $n$ variables, the exactness can be established.
This method is related to the Gumbel-max trick~\cite{gumbel1954statistical}, which has been applied in machine learning~\cite{huijben2022review}.

The Poisson functional representation supports a ``query operation'' whereby, given an input distribution $P$ over a space $\mathcal{U}$, it outputs a sample $\mathbf{U}_{P}$ drawn according to $P$. 
In communication settings~\cite{li2018strong, li2021unified}, this mechanism is used as follows: the encoder queries the representation using the prior distribution of the signal to obtain the signal to be transmitted, while the decoder queries it using the posterior distribution of the signal given the noisy observation to recover the message. 
The communication is successful if the two queries return the same sample.

We can generalize the Poisson functional representation by letting $\mathbf{U}_{P}(1),\ldots,\mathbf{U}_{P}(|\mathcal{U}|)\in\mathcal{U}$
be a random sequence formed by the elements of $\mathcal{U}$ sorted in ascending order of $Z_{u}/P(u)$:\footnote{Since $Z_u / P(u)$ are random, the sequence $\mathbf{U}_{P}(1),\ldots,\mathbf{U}_{P}(|\mathcal{U}|)$, sorted according to random values of $Z_{u}/P(u)$, is also random. More precisely, if we define a function $f:\mathbb{R}^{|\mathcal{U}|} \to \mathcal{U}^{|\mathcal{U}|}$ which maps $(z_u)_{u \in \mathcal{U}}$ to a sequence $u_1,\ldots,u_{|\mathcal{U}|}$ sorted in ascending order of $z_u / P(u)$, then we have $(\mathbf{U}_{P}(i))_i = f((Z_u)_u)$, i.e., $(\mathbf{U}_{P}(i))_i$ is the result of inputting random variables $(Z_u)_u$ into the function $f$.}
\[
\frac{Z_{\mathbf{U}_{P}(1)}}{P(\mathbf{U}_{P}(1))}\le\cdots\le\frac{Z_{\mathbf{U}_{P}(|\mathcal{U}|)}}{P(\mathbf{U}_{P}(|\mathcal{U}|))}.
\]
We break ties arbitrarily and treat $1/0=\infty$. 
In other words, we first generate $Z_u \sim \mathrm{Exp}(1)$ i.i.d. for $u \in \mathcal{U}$, and then find $\mathbf{U}_{P}(1) \in \mathcal{U}$ that gives the smallest $Z_u / P(u)$, find $\mathbf{U}_{P}(2) \in \mathcal{U} \backslash \{\mathbf{U}_{P}(1)\}$ that gives the smallest $Z_u / P(u)$ among the remaining elements, and so on.
This is similar to the mapped Poisson process in the generalized Poisson matching lemma \cite{li2021unified}, though unlike \cite{li2021unified},
$\mathbf{U}_{P}(1),\ldots,\mathbf{U}_{P}(|\mathcal{U}|)$ is not an
i.i.d. sequence following $P$. Write $\mathbf{U}_{P}^{-1}: \mathcal{U} \to [|\mathcal{U}|]$ for
the inverse function of $i \mapsto \mathbf{U}_{P}(i)$. 
The following is a
direct corollary of the generalized Poisson matching lemma~\cite{li2021unified}.

\myskip

\begin{lemma}[Generalized Poisson matching lemma~\cite{li2021unified}] 
\label{lem::GPML}
For distributions $P,Q$ over $\mathcal{U}$, 
we have the following for every $u\in \mathcal{U}$ with $P(u),Q(u)>0$:
\[
\mathbf{E}\left[\mathbf{U}_{Q}^{-1}(u)\,\Big|\,\mathbf{U}_{P} = u\right]\le\frac{P(u)}{Q(u)}+1.
\]
\end{lemma}
\myskip

The generalized Poisson matching lemma provides a bound on the mismatch between Poisson functional representations applied to different distributions. 
It enables analyses that require fewer (or no) uses of sub-codebooks and binning~\cite{li2021unified}, thereby reducing the number of error events and allowing the derivation of sharper ones-shot bounds. 
Note that Lemma~\ref{lem::GPML} is equivalent to stating that 
$\mathbf{E}[\mathbf{U}_{Q}^{-1}(\mathbf{U}_{P})\,|\,\mathbf{U}_{P}]\leq P(\mathbf{U}_{P}) / Q(\mathbf{U}_{P})+1$ holds almost surely.

We illustrate the use of the generalized Poisson matching lemma with channel coding as a toy example, where we derive a one-shot achievable bound that is similar to (but not exactly the same as) the one-shot bound for channel coding in~\cite{li2021unified}, and is a slightly weaker version of the dependence testing bound \cite{polyanskiy2010channel} (also see~\cite{hayashi2009information}). 
In channel coding, upon observing $M \sim \mathrm{Unif}[\mathsf{L}]$, the encoder produces $X$ and sends it through the channel $P_{Y|X}$. 
The decoder observes $Y$ and recovers $\hat{M}$. 
Let $\mathbf{U}$ be an i.i.d. exponential process over $\mathcal{U} = \mathcal{X} \times [\mathsf{L}]$, which serves as the ``random codebook''. 
Upon observing $M=m$, the encoder produces $\mathbf{U}_{P} \in \mathcal{X} \times [\mathsf{L}]$ where $P = P_X \times \delta_m$ is the ``encoding distribution'', and sends the $X$-component of $\mathbf{U}_{P}$ to the channel. 
Upon observing $Y=y$, the decoder computes $\mathbf{U}_{Q}$ where $Q = P_{X|Y}(\cdot | y) \times P_M$ is the ``decoding distribution'', and recovers $\hat{M}$, the $M$-component of $\mathbf{U}_{Q}$. 
The error probability is bounded as follows. 
\begin{align}
    & \mathbf{P} \big(M \neq \hat{M} \big)\nonumber \\
    &  \leq \mathbf{P} \big((X, M) \neq \mathbf{U}_{P_{X|Y}(\cdot | Y) \times P_M} \big)\nonumber\\
    & = \mathbf{E}\Big[
    \mathbf{P}\left( \mathbf{U}_{P_X \times \delta_m} \neq \mathbf{U}_{P_{X|Y}(\cdot | Y) \times P_M}\Big| \mathbf{U}_{P_X \times \delta_m} \Big) 
    \right] \nonumber
    \\
    &  =  \mathbf{E}\Big[
    \mathbf{P}\left( \mathbf{U}_{P_{X|Y}(\cdot | Y) \times P_M}^{-1} \big(\mathbf{U}_{P_X \times \delta_m} \big) > 1 
    \Big| \mathbf{U}_{P_X \times \delta_m} \right) 
    \Big] \nonumber \\
    &  \stackrel{(a)}{\leq} 
    \mathbf{E}\Big[\min\left\{
    \mathbf{E}\Big[
    \mathbf{U}_{P_{X|Y}(\cdot | Y) \times P_M}^{-1} \big(\mathbf{U}_{P_X \times \delta_m}\big) - 1 \Big| \mathbf{U}_{P_X \times \delta_m} 
    \Big]   , 1\right\}  
    \Big] \nonumber\\
    &  \stackrel{(b)}{\leq} \mathbf{E}\left[ \min\left\{
    P(\mathbf{U}_{P_X \times \delta_m})\big/ Q(\mathbf{U}_{P_X \times \delta_m}) 
    , 1\right\} \right] \nonumber \\
    &  = \mathbf{E}\left[ \min\left\{
    \mathsf{L}\cdot P_X(X)/P_{X|Y}(X|Y) 
    , 1\right\} \right] \nonumber \\
    &  = \mathbf{E}\left[ \min\left\{
    \mathsf{L}\cdot 2^{-\iota(X; Y)} 
    , 1\right\} \right], 
    \label{eq::PML_2}
\end{align}
where $(a)$ is by the Markov's inequality and $(b)$ is by the generalized Poisson matching lemma (Lemma~\ref{lem::GPML}). Note that we require the ``codebook'' $\mathbf{U}$, which is a public randomness available to the encoder and the decoder. Since $\mathbf{P} (M \neq \hat{M} ) = \mathbf{E}[\mathbf{P} (M \neq \hat{M} | \mathbf{U})]$, we can eliminate the need of public randomness without increasing the error probability by choosing a particular deterministic value $\mathbf{u}$ such that $\mathbf{P} (M \neq \hat{M} | \mathbf{U} = \mathbf{u}) \le \mathbf{E}[\mathbf{P} (M \neq \hat{M} | \mathbf{U})]$, and fixing the codebook to be $\mathbf{u}$.

\section{Exponential Process Refinement Lemma
\label{sec:expon}}

In this section, we design a new technique for proving one-shot achievability results over noisy multi-hop networks. 
The technique is named as the \emph{exponential process refinement lemma}, which is based on the Poisson matching lemma~\cite{li2021unified}.  
This lemma is useful for bounding the probability of error when the decoder attempts to recover multiple messages and/or sources via simultaneous (unique/non-unique) decoding. By having a simple bound that can be applied to general networks, we can derive coding theorems in a more systematic manner.

We first define a convenient tool as follows. 

\myskip
\begin{definition}[Refining a distribution by an exponential process]
\label{def:refine}
For a joint distribution $Q_{V,U}$
over $\mathcal{V}\times\mathcal{U}$, the refinement of $Q_{V,U}$ by $\mathbf{U}$, denoted as $Q_{V,U}^\mathbf{U}$, is a joint probability mass function 
\[
Q_{V,U}^\mathbf{U}(v,u):= 
\frac{Q_{V}(v)}{\mathbf{U}_{Q_{U|V}(\cdot|v)}^{-1}(u)\sum_{i=1}^{|\mathcal{U}|}i^{-1}}
\]
for all $(v, u)$ in the support of $Q_{V,U}$, 
where $Q_{V}$ is the
$V$-marginal of $Q_{V,U}$ and $Q_{U|V}$ is the conditional distribution of $U$ given $V$. 
When $V = \emptyset$, the above definition becomes
\[
Q_{U}^\mathbf{U}(u)=\frac{1}{\mathbf{U}_{Q_{U}}^{-1}(u)\sum_{i=1}^{|\mathcal{U}|}i^{-1}}.
\]
\end{definition}
\myskip

Note that the Poisson functional representation $\mathbf{U}_{Q_U} \in \mathcal{U}$ is a random variable which depends on the random exponential process $\mathbf{U}$, whereas the refinement $Q_{U}^\mathbf{U}(u)$ is a \emph{random distribution} over $\mathcal{U}$, i.e., it is a probability mass function over $\mathcal{U}$ whose probability values depend on $\mathbf{U}$.
The Poisson functional representation $\mathbf{U}_{Q_U}$, which only gives one value of $U$, is useful for unique decoding when the decoder only wants to recover a single value of the message or source. In comparison, the refinement $Q_{U}^\mathbf{U}(u)$ is for the \emph{soft decoding} of $U$, where the decoder does not output a single value, but rather a distribution indicating the likelihood of each value of the message or source.

Soft decoding is useful for \emph{non-unique decoding}, where the decoder wants to decode $V$ while utilizing the information in a correlated random variable $U$ without necessarily decoding $U$.\footnote{Non-unique decoding has been utilized in asymptotic analysis in network information theory, such as~\cite{nair2009capacity, chong2008han}.} The decoder can first obtain the refinement distribution $Q_{U}^{\mathbf{U}}$, and then compute the marginal distribution of $V$ in $Q_{U}^{\mathbf{U}} P_{V|U}$ (i.e., passing the distribution $Q_{U}^{\mathbf{U}}$ through the conditional distribution $P_{V|U}$) and use this marginal distribution to recover $V$ via the Poisson functional representation.
Using the soft decoding $Q_{U}^{\mathbf{U}}$, the decoder does not have to know $U$, and the decoding of $V$ can be successful as long as $Q_{U}^{\mathbf{U}}$ assigns a large enough probability to the correct $U$.

Loosely speaking, if the distribution $Q_{V,U}$ represents our ``prior distribution'' of $(V,U)$, then the refinement $Q_{V,U}^\mathbf{U}$ is our updated ``posterior distribution'' after taking the exponential process $\mathbf{U}$ into account. 
In multiterminal coding settings where a node decodes multiple random variables, the prior distribution of those random variables will be refined by multiple exponential processes. 
To keep track of the evolution of the ``posterior probability'' of the correct values of those random variables through the refinement process, we use the following lemma, called the \emph{exponential process refinement lemma}. 
Although its proof still relies on the Poisson matching lemma~\cite{li2021unified}, it significantly simplifies our analyses. 

\myskip
\begin{lemma}[Exponential Process Refinement Lemma]
\label{lemma::PML2}
For a distribution $P$ over $\mathcal{U}$ and a joint distribution
$Q_{V,U}$ over a finite $\mathcal{V}\times\mathcal{U}$, for every
$v\in\mathcal{V}$, we have, almost surely,
\begin{align*}
 \mathbf{E}\bigg[\frac{1}{Q_{V,U}^\mathbf{U}(v,\mathbf{U}_{P})}\bigg|\mathbf{U}_{P}\bigg] 
 \le \frac{\ln|\mathcal{U}|+1}{Q_{V}(v)}\left(\frac{P(\mathbf{U}_{P})}{Q_{U|V} (\mathbf{U}_{P}|v)}+1\right).
\end{align*}
\end{lemma}
\begin{IEEEproof}
We have 
\begin{align*} & \mathbf{E}\bigg[\frac{1}{Q_{V,U}^\mathbf{U}(v,\mathbf{U}_{P})}\,\bigg|\,\mathbf{U}_{P}\bigg] \\ 
& \stackrel{(a)}{=}   \mathbf{E}\bigg[  \frac{ \mathbf{U}_{Q_{U|V(\cdot | v)}}^{-1}(\mathbf{U}_{P})\sum_{i=1}^{|\mathcal{U}|}i^{-1}}{Q_{V}(v)}\Bigg|\mathbf{U}_P \bigg] \\ 
& \stackrel{(b)}{\leq }  \frac{\sum_{i=1}^{|\mathcal{U}|}i^{-1}}{Q_{V}(v)} \left( \frac{P(\mathbf{U}_{P})}{Q_{U|V}(\mathbf{U}_{P}|v)} + 1 \right) \\  
& \stackrel{(c)}{\leq }  \frac{\ln|\mathcal{U}|+1}{Q_{V}(v)}\left(\frac{P(\mathbf{U}_{P})}{Q_{U|V}(\mathbf{U}_{P}|v)}+1\right),
\end{align*} 
where $(a)$ is by Definition~\ref{def:refine}, $(b)$ is by Lemma~\ref{lem::GPML} and $(c)$ is by $\sum_{i=1}^n i^{-1} \leq \int_1^n x^{-1} dx + 1 = \ln n+1 $.  
\end{IEEEproof}

\mysmallskip

\section{Network Model}
\label{sec::net_model}

\begin{figure}[htpb]
	\centering
	\includegraphics[scale=0.25]{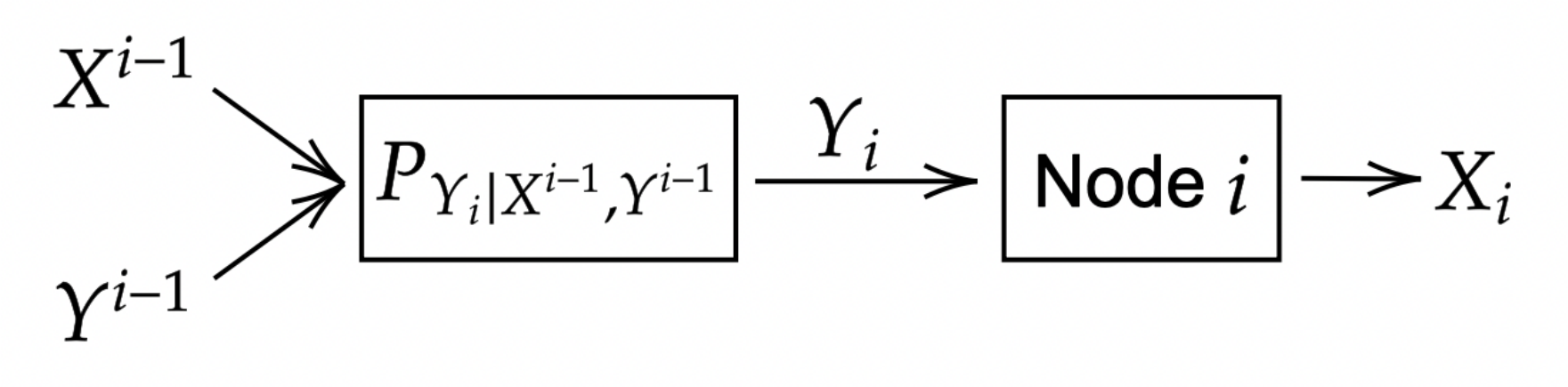}
	\caption{Acyclic discrete memoryless network.} 
	\label{fig::ADN_net}
\end{figure}

We describe a general $N$-node network model, which is the one-shot version of the \emph{acyclic discrete memoryless network (ADMN)}~\cite{lee2018unified}. 
There are $N$ nodes labelled $1,\ldots,N$. 
Node $i$ observes $Y_i \in \mathcal{Y}_i$ and produces $X_i \in \mathcal{X}_i$ (while we assume $\mathcal{X}_i,\mathcal{Y}_i$ are finite). 
Unlike conventional asymptotic settings (e.g. \cite{lee2018unified}), here $X_i$ is only \emph{one symbol}, instead of a sequence $(X_{i,1},\ldots,X_{i,n})$.
The transmission is performed sequentially, and each $Y_i$ is allowed to depend on all previous inputs and outputs (i.e., $X^{i-1},Y^{i-1}$) in a stochastic manner, as shown in Figure~\ref{fig::ADN_net}. 
Therefore, we can formally define an $N$\emph{-node acyclic discrete network (ADN)} as a collection of channels 
$(P_{Y_i | X^{i-1},Y^{i-1}})_{i\in [N]}$, where $P_{Y_i | X^{i-1},Y^{i-1}}$ is a conditional distribution from $(\prod_{j=1}^{i-1}\mathcal{X}_j) \times (\prod_{j=1}^{i-1}\mathcal{Y}_j)$ to $\mathcal{Y}_i$. 
In particular, $Y_1$ follows $P_{Y_1}$ and does not depend on any other random variable.
The asymptotic ADMN \cite{lee2018unified} can be considered as the $n$-fold ADN $(P^n_{Y_i | X^{i-1}, Y^{i-1}})_{i\in [N]}$, where $P^n_{Y_i | X^{i-1},Y^{i-1}}$ denotes the $n$-fold product conditional distribution (i.e., $n$ copies of a memoryless channel), and we take the blocklength $n \to \infty$.

We remark that, similar to the asymptotic unified random coding bound~\cite{lee2018unified}, the $X_i$'s and $Y_i$'s can represent sources, states, channel inputs, outputs and messages in source coding and channel coding settings. 
For example, for point-to-point channel coding, we take $Y_1$ to be the message, which the encoder (node $1$) encodes into the channel input $X_1$, which in turn is sent through the channel $P_{Y_2|X_1}$. 
The decoder (node $2$) observes $Y_2$ and outputs $X_2$, which is the decoded message. 
For lossless source coding, $Y_1$ is the source, $X_1=Y_2$ is the description by the encoder, and $X_2$ is the reconstruction.

\begin{figure}[htpb]
    \centering
    \includegraphics[scale=0.28]{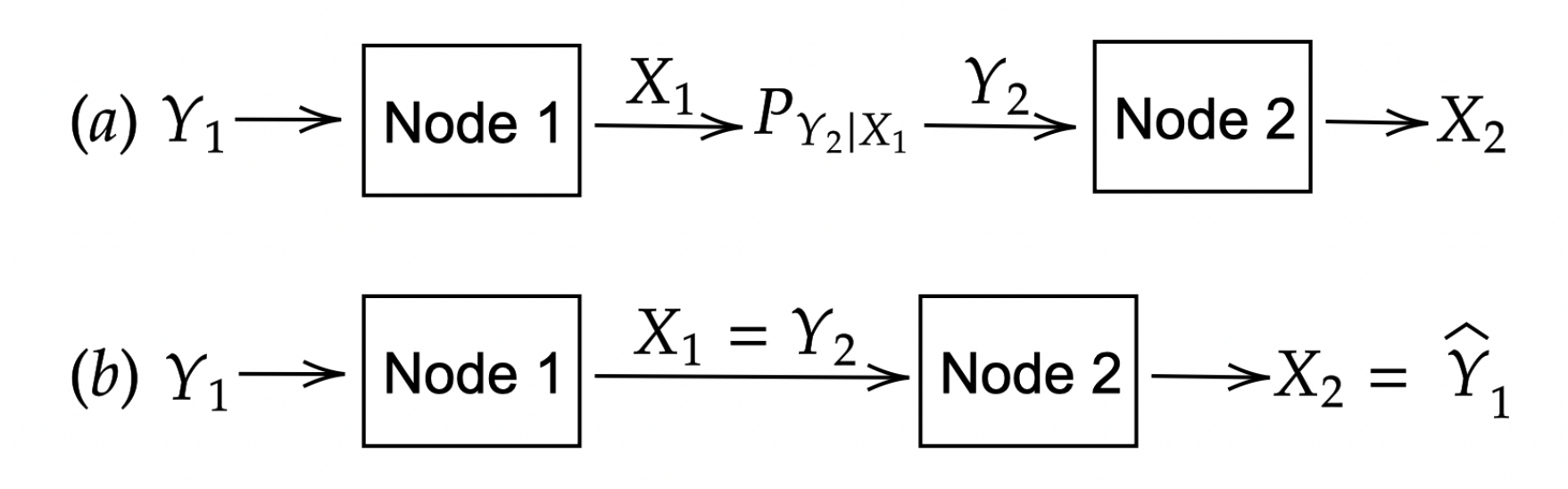}
    \caption{(a) Channel coding. (b) Source coding.} 
    \label{fig::channel_src}
\end{figure}
We give the definition of a coding scheme below.
\myskip
\begin{definition}
A \emph{deterministic coding scheme} consists of a sequence of encoding functions $(f_i)_{i\in [N]}$, where $f_i : \mathcal{Y}_i \to \mathcal{X}_i$.
For $i=1,\ldots,N$, the following operations are performed:
\begin{itemize}
    \item \textbf{Noisy channel.} The output $\tilde{Y}_i$ is generated conditional on $\tilde{X}^{i - 1},\tilde{Y}^{i-1}$ according to $P_{Y_i | X^{i-1},Y^{i-1}}$. 
    For $i=1$, $\tilde{Y}_1 \sim P_{Y_1}$  can be regarded as a source or a channel state. 

    \item \textbf{Node operation.} Node $i$ observes $\tilde{Y}_i$ and outputs $\tilde{X}_i = f_i(\tilde{Y}_i)$.
\end{itemize}
\end{definition}
\myskip

We sometimes allow an additional unlimited public randomness available to all nodes. 
\myskip
\begin{definition}
A \emph{public-randomness coding scheme} for the network consists of a pair $(P_W, (f_i)_{i\in [N]})$, where $P_W$ is the distribution of the public randomness $W \in \mathcal{W}$ available to all nodes and $f_i : \mathcal{Y}_i \times \mathcal{W} \to \mathcal{X}_i$ is the encoding function of node $i$ mapping its observation $Y_i$ and the public randomness $W$ to its output $X_i$.
The operations are as follows. 
First, generate $W \sim P_W$. 
For $i=1,\ldots,N$, generate $\tilde{Y}_i$ conditional on $\tilde{X}^{i-1},\tilde{Y}^{i-1}$ according to $P_{Y_i | X^{i-1},Y^{i-1}}$, 
and take $\tilde{X}_i = f_i(\tilde{Y}_i,W)$.
\end{definition}
\myskip

We do not impose any constraint on the public randomness $W$. 
In reality, to carry out a public-randomness coding scheme, the nodes share a common random seed to initialize their pseudorandom number generators before the scheme commences.

We use $\tilde{X}_i,\tilde{Y}_i$ to denote the actual random variables from the coding scheme. 
In contrast, $X_i,Y_i$ usually denote the random variables following an ideal distribution.
For example, in channel coding, the ideal distribution is $Y_1=X_2 \sim \mathrm{Unif}[\mathsf{L}]$ (i.e., the message is decoded without error), independent of $(X_1,Y_2) \sim P_{X_1}P_{Y_2|X_1}$. 
If we ensure that the actual $\tilde{X}^2,\tilde{Y}^2$ is ``close to'' the ideal $X^2,Y^2$, this would imply that $\tilde{Y}_1=\tilde{X}_2$ with high probability as well, giving a small error probability.

The goal (the ``achievability'') is to make the actual joint distribution $P_{\tilde{X}^N,\tilde{Y}^N}$ ``approximately as good as'' the ideal joint distribution $P_{X^N,Y^N}$. 
If we have an ``error set'' $\mathcal{E} \subseteq \big(\prod_{i=1}^N \mathcal{X}_i\big) \times \big(\prod_{i=1}^N \mathcal{Y}_i\big)$ that we do not want $(\tilde{X}^N,\tilde{Y}^N)$ to fall into (e.g., for channel coding, $\mathcal{E}$ is the set where $\tilde{Y}_1 \neq \tilde{X}_2$, i.e., an error occurs; for lossy source coding, $\mathcal{E}$ is the set where $d(\tilde{Y}_1, \tilde{X}_2)>\mathsf{D}$, i.e., the distortion exceeds the limit), we want 
\begin{align}
\mathbf{P}\big((\tilde{X}^N,\tilde{Y}^N) \in \mathcal{E} \big) \lesssim \mathbf{P}\big((X^N,Y^N) \in \mathcal{E} \big).\label{eq:errorset}
\end{align}
If $P_{\tilde{X}^N,\tilde{Y}^N}$ is close to $P_{X^N,Y^N}$ in total variation distance, i.e., 
\begin{align}
\delta_{\mathrm{TV}}\big(P_{X^N,Y^N},\, P_{\tilde{X}^N,\tilde{Y}^N} \big) \approx 0, 
\label{eq:error_tv}
\end{align}
then \eqref{eq:errorset} is guaranteed. 
For public-randomness coding schemes, in the next section, we will give a sufficient condition under which \eqref{eq:error_tv} can be achieved, which can be seen as a channel simulation~\cite{bennett2002entanglement, cuff2013distributed} or a coordination~\cite{cuff2010coordination} result. 
For deterministic coding schemes, since the node operations are deterministic, there might not be sufficient randomness to make $P_{\tilde{X}^N,\tilde{Y}^N}$ close to $P_{X^N,Y^N}$, and hence we use the  error bound in \eqref{eq:errorset} instead.

\section{Main Theorem for Acyclic Discrete Networks} 
\label{sec::main}

In this section, we present one-shot achievability results for ADN, using either a public-randomness coding scheme or a deterministic coding scheme.

Recall from Section~\ref{sec::net_model} that we have defined the acyclic discrete network (ADN) as an $N$-node network characterized by a collection of channels 
$(P_{Y_i | X^{i-1}, Y^{i-1}})_{i \in [N]}$. 
In addition to the network model, we also require a sequence $(a_{i,j})_j$ for node $i$ to specify which auxiliary random variables $U_j$ are decoded by node~$i$, and in what order. 
The main theorem stated as follows can be understood as a ``machine prover'', in the sense that with an ADN, a collection of auxiliary variables and a collection of decoding orders as input, this ``machine'' will provide a one-shot bound.

\myskip
\begin{theorem}
\label{thm::network_achievability}
Fix any ADN $(P_{Y_i | X^{i-1},Y^{i-1}})_{i\in [N]}$. 
For any collection of indices $(a_{i,j})_{i\in [N], j \in [d_i]}$ where $(a_{i,j})_{j \in [d_i]}$ is a sequence of distinct indices in $[i-1]$ for each $i$, any sequence $(d'_i)_{i\in [N]}$ with $0\le d'_i \le d_i$ and any collection of conditional distributions $(P_{U_i|Y_i, \overline{U}'_i}, P_{X_i|Y_i,U_i, \overline{U}'_i})_{i \in [N]}$ (where $\overline{U}_{i,\mathcal{S}} := (U_{a_{i,j}})_{j\in \mathcal{S}}$ for $\mathcal{S} \subseteq [d_i]$ and $\overline{U}'_{i}:=\overline{U}_{i,[d'_i]}$), which induces the joint distribution of $X^N,Y^N,U^N$ (the ``ideal distribution''), there exists a public-randomness coding scheme $(P_W, (f_i)_{i\in [N]})$ such that the joint distribution of $\tilde{X}^N, \tilde{Y}^N$ induced by the scheme (the ``actual distribution'') satisfies 
\begin{equation*}
\delta_{\mathrm{TV}}\big(P_{X^N,Y^N},\, P_{\tilde{X}^N,\tilde{Y}^N} \big)  \le
\mathbf{E}\bigg[\min\bigg\{
\sum_{i=1}^N 
\sum_{j=1}^{d'_i} 
B_{i,j},\, 1
\bigg\}
\bigg],
\end{equation*}
where 
\begin{align}
B_{i,j}  := &\gamma_{i,j} \prod_{k =j}^{d_i} \bigg(
  2^{-\iota
	(\overline{U}_{i,k}; \overline{U}_{i, [d_i] \backslash [j..k]} , Y_i) + \iota(\overline{U}_{i,k}; \overline{U}'_{a_{i,k}}, Y_{a_{i,k}}) }\nonumber \\
    & \qquad \qquad  \qquad + \mathbf{1}\{k \! >\! j\}\bigg) \label{eq::beta}
\end{align} 
such that\footnote{Note the logarithmic terms $\gamma_{i,j}$ do not affect the first and second order results.} \begin{align*}
&\gamma_{i,j} := \prod_{k=j+1}^{d_i}
\Big( \ln|\mathcal{U}_{a_{i,k}}| + 1\Big).
\end{align*} 

\end{theorem}
\myskip

We also have the following bound for deterministic coding schemes, which can be obtained by fixing a particular choice of the public randomness. 

\myskip
\begin{theorem}\label{thm:det}
Fix any ADN $(P_{Y_i | X^{i-1},Y^{i-1}})_{i\in [N]}$.
For any $(a_{i,j})_{i\in [N], j \in [d_i]}$, $(d'_i)_{i\in [N]}$, $(P_{U_i|Y_i, \overline{U}'_i}, P_{X_i|Y_i,U_i, \overline{U}'_i})_{i \in [N]}$ as defined in Theorem~\ref{thm::network_achievability}, 
which induce the joint distribution of $X^N,Y^N,U^N$, and any set $\mathcal{E} \subseteq \big(\prod_{i=1}^N \mathcal{X}_i\big) \times \big(\prod_{i=1}^N \mathcal{Y}_i\big)$, there is a deterministic coding scheme $(f_i)_{i\in [N]}$ such that $\tilde{X}^N,\tilde{Y}^N$ induced by the scheme satisfy
\begin{align}
&\mathbf{P}\big((\tilde{X}^N,\tilde{Y}^N) \in \mathcal{E} \big) \nonumber \\
&\le \mathbf{E}\bigg[\min\bigg\{
\mathbf{1}\big\{(X^N,Y^N) \in \mathcal{E} \big\} + \sum_{i=1}^N 
\sum_{j=1}^{d'_i} 
B_{i,j},\, 1
\bigg\}
\bigg], 
\label{eq:error_bound_det}
\end{align}
where $B_{i,j}$ is defined in Theorem~\ref{thm::network_achievability}. 
\end{theorem}
\myskip

We briefly explain Theorem~\ref{thm::network_achievability} and Theorem~\ref{thm:det} as follows.
The sequences $(a_{i,j})_j$ control which auxiliaries $U_j$ node $i$ decodes and in which order. 
Node $i$ uniquely decodes $\overline{U}'_i = $ $(U_{a_{i,j}})_{j \in [d'_i]}$ while utilizing $(U_{a_{i,j}})_{j \in [d'_i+1 .. d_i]}$ by non-unique decoding via the exponential process refinement (Definition~\ref{def:refine}). 
For brevity, we say ``the \emph{decoding order} of node $i$ is $\overline{U}_{i,1}, \ldots,$ $\overline{U}_{i,d'_i}, \overline{U}_{i,d'_i+1}?,\ldots,\overline{U}_{i,d_i}?$'' where ``?'' means the random variable is only used in non-unique decoding. 
Node $i$ decodes $\overline{U}'_i$, creates its own $U_i$ by using the Poisson functional representation on $P_{U_i|Y_i, \overline{U}'_i}$, and generates $X_i$ from $P_{X_i|Y_i,U_i, \overline{U}'_i}$. 

For example, for channel coding~\eqref{eq::PML_2} where the setting is also shown in Fig.~\ref{fig::channel_src}~(a), we can consider it as an ADN and utilize our theorem as follows: take $Y_1 = M \sim \mathrm{Unif}[\mathsf{L}]$ to be the message as the input of node~1 (encoder), $X_1 = X$ to be the channel input; $P_{Y_2|X_1} = P_{Y|X}$ to be the noisy channel, $Y_2 = Y$ to be the channel output observed by node~2 (decoder); and $X_2 = M$ to be the ideal output of node~2.  
The auxiliary variable of node~1 is $U_1 = (X, M)$.  
The decoding order of node~2 is ``$U_1$'' (i.e., it only wants $U_1$), hence in this case $d_2' = d_2 = 1$ and $a_{2, 1} = 1$. 
Since node~2 has decoded $U_1$, $X_2$ is allowed to depend on $U_1 = (X, M)$, and therefore the choice $X_2 = M$ is valid in the ideal scenario, though in the actual scenario the decoder's output $\tilde{X}_2$ is not exactly $M$. Take the error set to be $\mathcal{E} = \{(x^2,y^2):\, y_1 \neq x_2\}$, i.e., it is the set where the decoder's output does not match the message. We can bound the error probability $\mathbf{P}((\tilde{X}^2,\tilde{Y}^2) \in \mathcal{E}) = \mathbf{P}(\tilde{Y}_1 \neq \tilde{X}_2)$ by Theorem~\ref{thm:det}, which recovers the bound in~\eqref{eq::PML_2}. 
More examples will be given later, which will hopefully clarify the roles of various quantities in Theorem~\ref{thm::network_achievability} and Theorem~\ref{thm:det}.

Theorem~\ref{thm::network_achievability} implies the following result for the asymptotic ADMN \cite{lee2018unified} by directly applying the law of large numbers.

\myskip

\begin{corollary}\label{cor:admn}
Fix any ADN $(P_{Y_i | X^{i-1},Y^{i-1}})_{i\in [N]}$.
Fix any $(a_{i,j})_{i\in [N], j \in [d_i]}$, $(d'_i)_{i\in [N]}$, $(P_{U_i|Y_i, \overline{U}'_i}, P_{X_i|Y_i,U_i, \overline{U}'_i})_{i \in [N]}$ as defined in Theorem~\ref{thm::network_achievability}, 
which induces the joint distribution of $X^N,Y^N,U^N$. If for every $i \in [N]$, $j \in [d'_i]$, 
\begin{align*}
& I(\overline{U}_{i,j}; \overline{U}_{i, [d_i] \backslash \{j\}} , Y_i) - I(\overline{U}_{i,j}; \overline{U}'_{a_{i,j}}, Y_{a_{i,j}})  > \\
&\qquad \sum_{k=j+1}^{d_i} \bigg( \max \big\{ I(\overline{U}_{i,k}; \overline{U}'_{a_{i,k}}, Y_{a_{i,k}}) \\
&\qquad \qquad \qquad \qquad - I(\overline{U}_{i,k}; \overline{U}_{i, [d_i] \backslash [j..k]} , Y_i),\,0 \big\}\bigg),
\end{align*}
then there is a sequence of public-randomness coding schemes (indexed by $n$) for the $n$-fold ADN $(P^n_{Y_i | X^{i-1},Y^{i-1}})_{i\in [N]}$ such that the induced $\tilde{X}^{N,n},\tilde{Y}^{N,n}$ (write $\tilde{X}^{N,n}=(\tilde{X}_{i,j})_{i\in [N], j \in [n]}$) satisfy
\begin{equation}
\lim_{n \to \infty} \delta_{\mathrm{TV}}\big(P^n_{X^{N},Y^{N}},\, P_{\tilde{X}^{N,n},\tilde{Y}^{N,n}} \big) = 0.
\end{equation}
\end{corollary}

\medskip
While this result is not as strong as the general asymptotic result in \cite{lee2018unified}, a one-shot analogue of \cite{lee2018unified} will likely be significantly more complicated than Theorem~\ref{thm::network_achievability}. 
We choose to present Theorem~\ref{thm::network_achievability} since it is simple but already general and powerful enough to give a wide range of tight one-shot results.

\section{One-shot Relay Channel}
\label{sec::relay}

\begin{figure*}[htpb]
\centering
\includegraphics[scale=0.36]{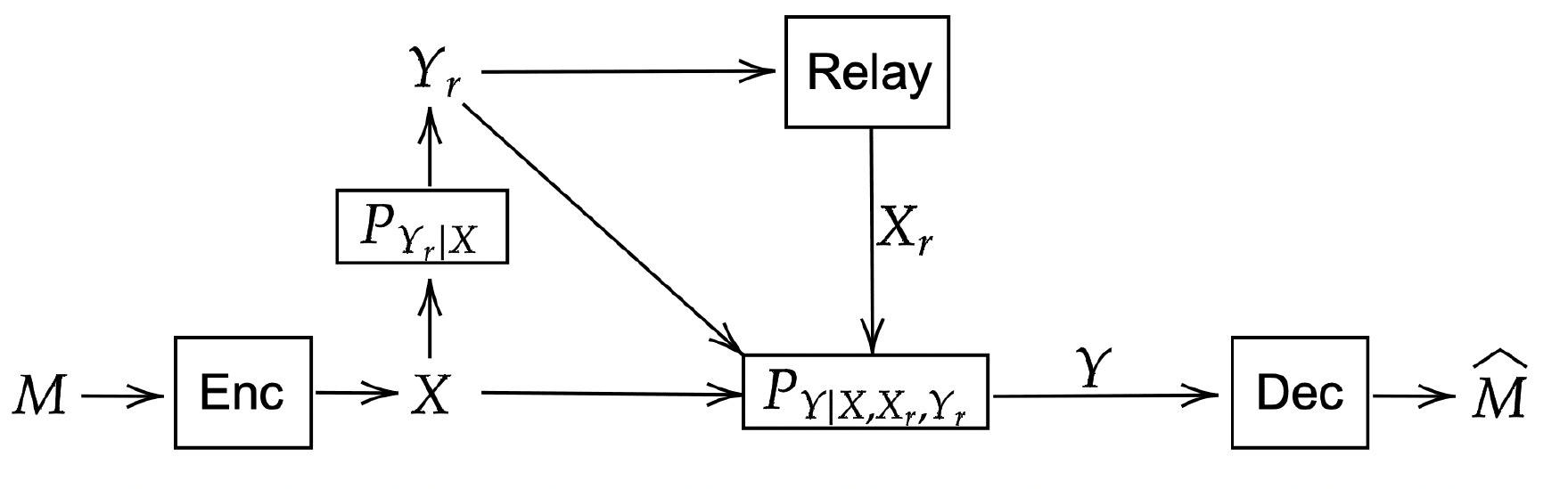}
\caption{One-shot relay channel setting.} \label{fig::setup_relay_one_shot}
\end{figure*}

To explain our scheme, we first discuss a \emph{one-shot relay channel} in Figure~\ref{fig::setup_relay_one_shot}. 
An encoder observes $M \sim \mathrm{Unif}[\mathsf{L}]$ and outputs $X$, which is passed through the channel $P_{Y_{\mathrm{r}}|X}$.
The relay observes $Y_{\mathrm{r}}$ and outputs $X_{\mathrm{r}}$.
Then $(X,X_{\mathrm{r}},Y_{\mathrm{r}})$ is passed through the channel $P_{Y|X,X_{\mathrm{r}},Y_{\mathrm{r}}}$. 
The decoder observes $Y$ and recovers $\hat{M}$.
For generality, we allow $Y$ to depend on all of $X,X_{\mathrm{r}},Y_{\mathrm{r}}$, and $X_{\mathrm{r}}$ may interfere with $(X,Y_{\mathrm{r}})$, 
which can happen if the relay outputs $X_{\mathrm{r}}$ instantaneously or the channel has a long memory, or it is a storage device. 
It is a one-shot version of the \emph{relay-without-delay} and \emph{relay-with-unlimited-look-ahead}~\cite{el2005relay, el2007relay}, and is an  ADN by taking $Y_1=M$, $X_1=X$, $Y_2=Y_{\mathrm{r}}$, $X_2=X_{\mathrm{r}}$, $Y_3=Y$, and $X_3=M$ (in the ideal distributions).

In case if $Y=(Y',Y'')$ consists of two components and the channel $P_{Y|X,X_{\mathrm{r}},Y_{\mathrm{r}}}=P_{Y'|X,Y_{\mathrm{r}}}P_{Y''|X_{\mathrm{r}}}$ can be decomposed into two orthogonal components (so $X_{\mathrm{r}}$ does not interfere with $(X,Y_{\mathrm{r}})$), this becomes the one-shot version of the \emph{primitive relay channel}~\cite{el2011network, kim2007coding, mondelli2019new, el2021achievable, el2022strengthened} since the $n$-fold version of this ADN (with $n\to \infty$) is precisely the asymptotic primitive relay channel.
However, the $n$-fold version of the ADN in Figure~\ref{fig::setup_relay_one_shot} in general is not the conventional relay channel~\cite{el2011network, van1971three, cover1979capacity} (it is the relay-with-unlimited-look-ahead~\cite{el2005relay, el2007relay} instead). 
The conventional relay channel, due to its causal assumption that the relay can only look at past $Y_{\mathrm{r},t}$'s, has no one-shot counterpart.

We use the following corollary of Theorem \ref{thm:det} to demonstrate the use of the exponential process refinement lemma (Lemma~\ref{lemma::PML2}).

\medskip

\begin{corollary}
\label{cor::one_shot_relay_eg}
For any $P_{X}$, $P_{U|Y_{\mathrm{r}}}$, function $x_{\mathrm{r}}(y_{\mathrm{r}},u)$, 
there is a deterministic coding scheme for the one-shot relay channel such that 
the error probability satisfies
\begin{equation} 
\label{eq::one_shot_relay}
      P_e\leq \mathbf{E} \Big[
    \min\big\{
    \gamma \mathsf{L} 
     2^{-\iota (X;U,Y)}
     \big(2^{-\iota (U;Y) + \iota (U;Y_{\mathrm{r}}) } +1 \big) , 1
    \big\}
    \Big],
\end{equation} 
where $(X,Y_{\mathrm{r}},U, X_{\mathrm{r}},Y) \sim$$P_{X} P_{Y_{\mathrm{r}}|X} P_{U|Y_{\mathrm{r}}} \delta_{x_{\mathrm{r}}(Y_{\mathrm{r}},U)} P_{Y|X,Y_{\mathrm{r}},X_{\mathrm{r}}}$, and $\gamma :=  \ln |\mathcal{U}| + 1$. 
\end{corollary}

\medskip

\begin{IEEEproof}
For the sake of demonstration, we first give a detailed proof via the exponential process refinement lemma without invoking Theorem \ref{thm:det}.
Let $U_1 := (X,M)$, $U_2 := U$. 
Let $\mathbf{U}_1$, $\mathbf{U}_2$ be two independent exponential processes, which serve as the ``random codebooks''. 
The encoder (node 1) uses the Poisson functional representation~\eqref{eq:pfr} to compute $U_1 = (\mathbf{U}_1)_{P_{U_1} \times  \delta_M }$ and outputs $X$-component of $U_1$. 
The relay (node 2) computes 
$U_2 = (\mathbf{U}_2)_{P_{U_2|Y_{\mathrm{r}}}(\cdot| Y_{\mathrm{r}})}$ and outputs $X_{\mathrm{r}}=x_{\mathrm{r}}(Y_{\mathrm{r}}, U_2)$. 
Note that $X,Y_{\mathrm{r}},U_2, X_{\mathrm{r}},Y$ follow the ideal distribution in the corollary due to the property of Poisson functional representation, and hence we write $X_{\mathrm{r}}$ instead of $\tilde{X}_{\mathrm{r}}$.
The decoder (node 3) observes $Y$, and performs the following steps. 

\begin{enumerate}
\item Refine $P_{U_2|Y}(\cdot|Y)$ (written as $P_{U_2|Y}$ for brevity) to ${Q}_{U_2} := P_{U_2|Y}^{\mathbf{U}_2}$ using Definition~\ref{def:refine}. 
By the exponential process refinement lemma (Lemma~\ref{lemma::PML2}, with $V = \emptyset$), 
\begin{align*}
     & \mathbf{E}
     \bigg[\frac{1}{{Q}_{U_2}(U_2)} \bigg| \, U_2,Y,Y_{\mathrm{r}}\bigg]
     \leq  \\
     & \qquad \qquad (\ln |\mathcal{U}_2|+1)
    \left(
    \frac{P_{U_2|Y_{\mathrm{r}}}(U_2)}{P_{U_2|Y}(U_2) } + 1
    \right).
\end{align*}

\item Compute the joint distribution ${Q}_{U_2} P_{U_1 | U_2,Y}$ over $\mathcal{U}_1 \times \mathcal{U}_2$, 
	the semidirect product between ${Q}_{U_2}$ and $P_{U_1 | U_2,Y}(\cdot | \cdot , Y)$. 
 Let its $U_1$-marginal be $\tilde{Q}_{U_1}$.

\item Let $\tilde{U}_1 = (\mathbf{U}_1)_{\tilde{Q}_{U_1}\times P_M}$, and output its $M$-component. 
\end{enumerate}
Letting $A :=( X,Y_{\mathrm{r}},U_2, X_{\mathrm{r}},Y ,M)$ and $\gamma := \ln|\mathcal{U}_2| + 1$, we have
\begin{align*}
    &\mathbf{P} (
    \tilde{U}_1 \neq U_1 \, | \, A
    ) \\
    & \stackrel{(a)}{\leq} \mathbf{E}\left[
    \min\left\{
      \frac{P_{U_1}(U_1)  \delta_M(M)}{ 
    P_{U_1 | U_2,Y}(U_1 | U_2,Y)
     {Q} _{U_2}(U_2)  P_M(M) } , 1
    \right\} \,\bigg|\, A   \right]  \\ 
    & \stackrel{(b)}{=} \mathbf{E}\left[
    \min\left\{
      \mathsf{L}\frac{P_{U_1}(U_1)}{ 
    P_{U_1 | U_2,Y}(U_1 | U_2,Y)
     {Q} _{U_2}(U_2)} , 1
    \right\} \,\bigg|\, A   \right]  \\ 
    & \stackrel{(c)}{\leq} 
    \min\bigg\{\mathsf{L}
      \frac{ P_{U_1}(U_1)}{ P_{U_1|U_2,Y} (U_1|U_2,Y)  } \gamma        \bigg(
    \frac{P_{U_2|Y_{\mathrm{r}}}(U_2)}{P_{U_2|Y}(U_2) } + 1
    \bigg)  , 1
    \bigg\}     \\
    & = 
    \min\big\{
    \gamma \mathsf{L}
     2^{-\iota (X;U_2,Y)}
     \big(2^{-\iota (U_2;Y) + \iota (U_2;Y_{\mathrm{r}}) } +1 \big) , 1
    \big\},
\end{align*}
where $(a)$ is by the generalized Poisson matching lemma~\cite{li2021unified} (Lemma \ref{lem::GPML}), $(b)$ is by $\delta_M(M)=1$ and $P_M(M)=1/\mathsf{L}$, and $(c)$ is by step 1) and Jensen's inequality. Taking expectation over $A$ gives the desired error bound. Although the codebooks $\mathbf{U}_1$, $\mathbf{U}_2$ are random (so this is a public-randomness scheme), we can convert it to a deterministic scheme by fixing one particular choice $(\mathbf{u}_1,\mathbf{u}_2)$ that satisfies the error bound.

\mysmallskip

Alternatively, Theorem \ref{thm:det} allows us to derive bounds for general acyclic discrete networks in a systematic manner, without going through the above arguments for every specific ADN. To prove Corollary \ref{cor::one_shot_relay_eg}, we can invoke Theorem~\ref{thm:det} on the ADN with nodes $1,2,3$, with inputs $Y_i$'s, outputs $X_i$'s, auxiliaries $U_i$'s and the terms $B_{i,j}$'s as follows:
\begin{enumerate}
    \item Node $1$ has input $Y_1 = M$, output $X_1 = X$ and auxiliary $U_1 = (X,M)$. 
    
    \item Node $2$ has input $Y_2 = Y_r$, output $X_2 = X_r$ and auxiliary $U_2 = U$.  
    
    \item Node $3$ has input $Y_3 = Y$, output $X_3 = M$ and decodes with the order ``$U_1, U_2?$''. 
    Applying Theorem~\ref{thm:det} (note that $d_3 =2$ and $d_3' =1$), we have 
    \begin{align*}
        B_{3,1} = (\ln|\mathcal{U}_2| + 1) 
        \mathsf{L} 2^{-\iota (X;U_2,Y)}
         \big(2^{-\iota (U_2;Y) + \iota (U_2;Y_{\mathrm{r}}) } +1 \big),
    \end{align*}
    and hence we obtain the bound \eqref{eq::one_shot_relay} by invoking Theorem \ref{thm:det}.

\end{enumerate} 
\end{IEEEproof}

\medskip

Corollary~\ref{cor::one_shot_relay_eg} yields the following asymptotic achievable rate: 
\begin{align*}
    &R\leq I(X;U, Y) - \max\big\{ I(U; Y_{\mathrm{r}}) - I(U; Y) ,\, 0\big\}
\end{align*}
for some $P_{U|Y_{\mathrm{r}}}$ and function $x_{\mathrm{r}}(y_{\mathrm{r}},u_2)$. 

\mysmallskip

\begin{figure*}[htpb]
	\centering
	\includegraphics[scale=0.37]{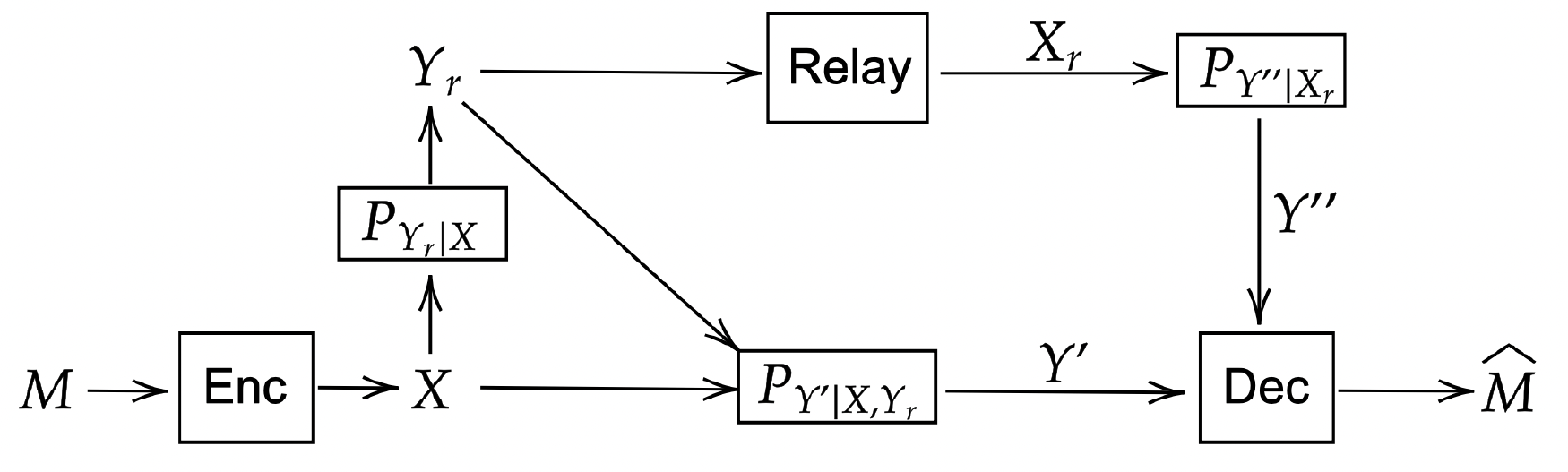}
	\caption{One-shot primitive relay channel setting. } 
    \label{fig::primitive_relay}
\end{figure*}

We also consider a one-shot primitive relay channel (as shown in Figure~\ref{fig::primitive_relay}), where $P_{Y|X,X_{\mathrm{r}}, Y_{\mathrm{r}}}=P_{Y'|X, Y_{\mathrm{r}}} P_{Y''|X_{\mathrm{r}}}$ can be decomposed into two orthogonal components. 
Consider $(X,Y_{\mathrm{r}},Y')$ independent of $(X_{\mathrm{r}},Y'')$ in the ideal distribution and take $U = (U',X_{\mathrm{r}})$ where $U'$ follows $P_{U'|Y_{\mathrm{r}}}$, Corollary~\ref{cor::one_shot_relay_eg} specializes to the following Corollary~\ref{cor::one_shot_primrelay_eg}.

\mysmallskip

\begin{corollary}
\label{cor::one_shot_primrelay_eg}
For any $P_{X}$, $P_{X_{\mathrm{r}}}$, $P_{U'|Y_{\mathrm{r}}}$, there is a deterministic coding scheme for the one-shot primitive relay channel with $M\sim\mathrm{Unif}[\mathsf{L}]$ such that the error probability satisfies
\begin{align*} 
    P_e\leq \mathbf{E} \Big[\!
    \min\!\Big\{
    \gamma \mathsf{L} 
    2^{-\iota (X;U',Y')}
    \big(2^{- \iota(X_{\mathrm{r}};Y'') + \iota (U';Y_{\mathrm{r}}|Y') } \! +\! 1 \big) , 1
    \!\Big\}\!
    \Big]\! ,
\end{align*}
where $(X,Y_{\mathrm{r}},U', Y') \sim P_{X} P_{Y_{\mathrm{r}}|X} P_{U'|Y_{\mathrm{r}}}P_{Y'|X,Y_{\mathrm{r}}}$ is independent of $(X_{\mathrm{r}},Y'') \sim P_{X_{\mathrm{r}}} P_{Y''|X_{\mathrm{r}}}$, and $\gamma :=  \ln (|\mathcal{U}'||\mathcal{X}_{\mathrm{r}}|) + 1$.
\end{corollary}

\mysmallskip

This gives the asymptotic achievable rate $R \le I(X;U',Y') -\max\{ I(U';Y_{\mathrm{r}}|Y') - C_{\mathrm{r}} ,\, 0 \}$ where $C_{\mathrm{r}} = \max_{P_{X_{\mathrm{r}}}}I(X_{\mathrm{r}};Y'')$ is the capacity of the channel $P_{Y''|X_{\mathrm{r}}}$. 
It implies the compress-and-forward bound~\cite{kim2007coding}, which is the maximum of $I(X;U',Y')$ subject to the constraint $C_{\mathrm{r}} \ge I(U';Y_{\mathrm{r}}|Y')$ (where the random variables are distributed as in Corollary~\ref{cor::one_shot_primrelay_eg}). Hence, Corollary~\ref{cor::one_shot_primrelay_eg} can be treated as a one-shot compress-and-forward bound.

\subsection{Partial-Decode-and-Forward Bound}

We extend Corollary~\ref{cor::one_shot_relay_eg} to allow partial decoding of the message~\cite{cover1979capacity,kim2007coding,el2007relay}.
To this end, we split the message and encoder into two. 
The message $M\sim \mathrm{Unif}[\mathsf{L}]$ is split into $M_1\sim \mathrm{Unif}[\mathsf{J}]$ and $M_2\sim \mathrm{Unif}[\mathsf{L}/\mathsf{J}]$ (assume $\mathsf{J}$ is a factor of $\mathsf{L}$).
The encoder controls two nodes (node $1$ and $2$), where node $1$ observes $Y_1 = M_1$, outputs $X_1=V$, and has an auxiliary $U_1=(M_1,V)$; node $2$ observes $Y_2=(M_1,M_2,V)$, outputs $X_2=X$, and has an auxiliary $U_2=(M_1,M_2,X)$.
The relay (node $3$) observes $Y_3 = Y_r$, decodes $U_1$, outputs $X_3 = X_r$, and has an auxiliary $U_3=(M_1,U)$.
The decoder (node $4$) observes $Y_4=Y$ and uses the decoding order ``$U_2,U_3?,U_1?$''. 

\myskip
\begin{corollary}
\label{cor::pdcf}
Fix any $P_{X, V}$, $P_{U|Y_{\mathrm{r}}, V}$, function $x_{\mathrm{r}}(y_{\mathrm{r}},u,v)$, and $\mathsf{J}$ which is a factor of $\mathsf{L}$.
There exists a deterministic coding scheme for the one-shot relay channel with
\begin{align*}
P_{e} & \leq\mathbf{E}\Big[\min\Big\{\mathsf{J}2^{-\iota(V;Y_{\mathrm{r}})}+\gamma\mathsf{L}\mathsf{J}^{-1}2^{-\iota(X;U,Y|V)} \\
&\;\;\;\;\;\;\;\;\; \cdot\big(2^{-\iota(U;V,Y)+\iota(U;V,Y_{\mathrm{r}})}+1\big)\big(\mathsf{J}2^{-\iota(V;Y)}+1\big),1\Big\}\Big],
\end{align*}
where $(X,V,Y_{\mathrm{r}},U, X_{\mathrm{r}},Y) \sim P_{X, V} P_{Y_{\mathrm{r}}|X,V} P_{U|Y_{\mathrm{r}}, V} \delta_{x_{\mathrm{r}}(Y_{\mathrm{r}},U, V)}  
P_{Y|X,Y_{\mathrm{r}},X_{\mathrm{r}}}$ and $\gamma :=  (\ln (\mathsf{J}|\mathcal{U}|) + 1) ( \ln(\mathsf{J}|\mathcal{V}|) +1)$.
\end{corollary}

\myskip

Applying the law of large numbers to Corollary~\ref{cor::pdcf}, and Fourier-Motzkin elimination (using the PSITIP software~\cite{li2023automated}), we obtain the following asymptotic achievable rate:
\[
\min\!\left\{\!\!\! \begin{array}{l}
I(V;Y)+I(U,Y;X|V),\\
I(V;Y_{\mathrm{r}})+I(U,Y;X|V),\\
I(V,U;Y)+I(U,Y;X|V)-I(U;Y_{\mathrm{r}}|V),\\
I(V;Y_{\mathrm{r}})+I(U;Y|V)+I(U,Y;X|V)-I(U;Y_{\mathrm{r}}|V)
\end{array}\!\!\!\right\}\! ,
\]
where $(X,V,Y_{\mathrm{r}},U, X_{\mathrm{r}},Y) \sim P_{X, V} P_{Y_{\mathrm{r}}|X,V} P_{U|Y_{\mathrm{r}}, V} \delta_{x_{\mathrm{r}}(Y_{\mathrm{r}},U, V)}  \linebreak[1] P_{Y|X,Y_{\mathrm{r}},X_{\mathrm{r}}}$, 
subject to the constraint $I(U;Y_{\mathrm{r}}|V)\le I(U;Y|V)+I(U,Y;X|V)$.

Taking $x_\mathrm{r}(y_\mathrm{r},(v',x'_\mathrm{r})) = x'_\mathrm{r}$, $U=\emptyset$, $V $ $=(V',X'_\mathrm{r})$, it gives an achievable rate $\min\{I(X,X_{\mathrm{r}};Y),\,I(V';Y_{\mathrm{r}})+I(X;  Y|X_{\mathrm{r}},V')\}$, recovering the partial noncausal decode-forward bound for relay-with-unlimited-look-ahead~\cite[Prop. 3]{el2007relay}.

Specializing to the primitive relay channel, and again substituting $U=\emptyset$, $V=(V',X'_\mathrm{r})$, $x_\mathrm{r}(y_\mathrm{r},(v',x'_\mathrm{r}))=x'_\mathrm{r}$, we have
\begin{align*}
P_{e} & \leq\mathbf{E}\Big[\min\Big\{\mathsf{J}2^{-\iota(V';Y_{\mathrm{r}})}+2\gamma\mathsf{L}\mathsf{J}^{-1}2^{-\iota(X;Y'|V')} \\
& \;\;\;\;\;\;\;\;\; \cdot\big(\mathsf{J}2^{-\iota(V';Y')-\iota(X_{\mathrm{r}};Y'')}+1\big),1\Big\}\Big],
\end{align*}
where $\gamma :=  (\ln \mathsf{J} + 1) ( \ln(\mathsf{J}|\mathcal{V}'||\mathcal{X}_\mathrm{r}|) +1)$ and $(X,V',Y_{\mathrm{r}},Y') \sim P_{X,V'} P_{Y_{\mathrm{r}}|X} P_{Y'|X,Y_{\mathrm{r}}}$ is independent of $(X_{\mathrm{r}},Y'') \sim P_{X_{\mathrm{r}}} P_{Y''|X_{\mathrm{r}}}$. 
It gives the asymptotic rate $\min\{I(V';Y_{\mathrm{r}})+I(X;Y|V'),\,I(X;Y)+C_{\mathrm{r}}\}$ and recovers the partial decode-forward lower bound for primitive relay channels~\cite{cover1979capacity,kim2007coding}.
One-shot versions of other asymptotic bounds for primitive relay channels (e.g., \cite{mondelli2019new,el2021achievable}) are left for future studies.

\section{Cascade multiterminal source coding with computing} 
\label{sec::cascade}

We consider the cascade multiterminal source coding problem~\cite{cuff2009cascade} (which is also called the \emph{cascade coding for computing} in~\cite[Section 21.4]{el2011network}).
It is similar to the traditional multiterminal source coding problem introduced by Berger and Tung~\cite{berger1978multiterminal, tung1978multiterminal}, where two information sources are encoded in a distributed fashion with loss, though the communication between encoders replaces one of the direct channels to the decoder in the cascade case. 
It can include different variations, e.g., the decoder desires to estimate both $X$ and $Y$, $X$ only, $Y$ only and some functions of both. 
It is tightly related to real problems where it is required to pass messages to neighbors in order to compute functions of data, e.g., distributed data collection, aggregating measurements in sensor networks, interactive coding for computing and distributed lossy averaging (see~\cite{cuff2009cascade} and references therein).

The asymptotic rate-distortion region for the general cascade multiterminal source coding problem is unknown, even for the case where $X$ and $Y$ are independent. 
We study the one-shot setting of this problem, which has not been discussed in literature to the best of our knowledge. 
We provide a novel one-shot bound on the cascade multiterminal source coding problem, and show that our one-shot achievability result recovers the best known asymptotic inner bound, i.e., the \emph{local-computing-and-forwarding inner bound}~\cite{cuff2009cascade} (which in turn recovers various other existing bounds as special cases, also see~\cite[Section 21.4]{el2011network} for a detailed discussion). 

\begin{figure}[htpb]
    \centering
    \includegraphics[scale=0.28]{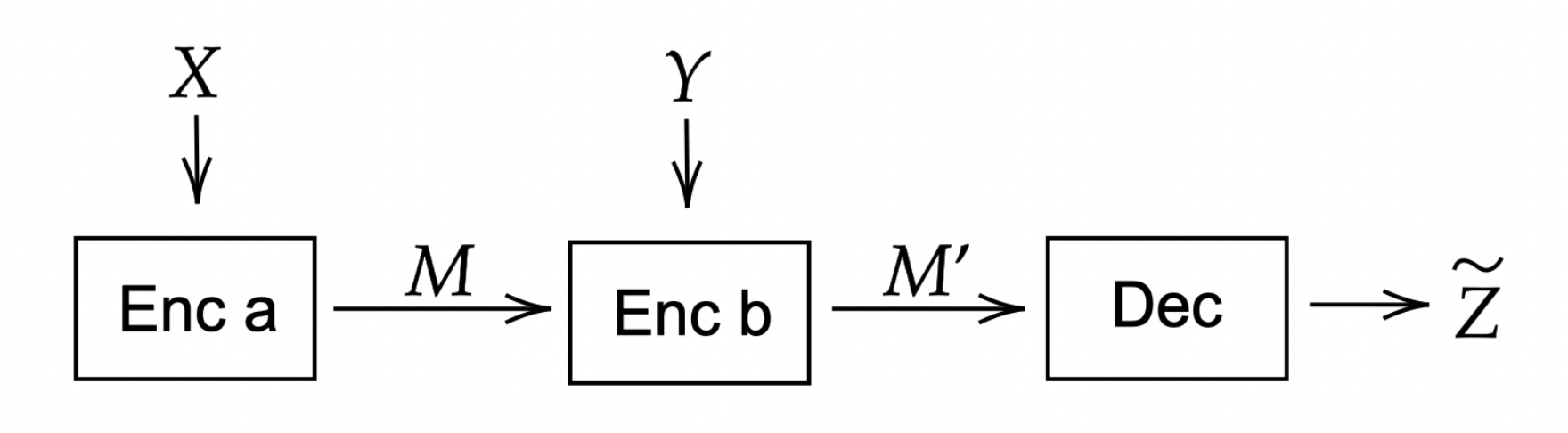}
    \caption{One-shot cascade multiterminal source coding setting. } 
    \label{fig::cascade_asym}
\end{figure}

The one-shot cascade multiterminal source coding problem is described as follows (see Figure~\ref{fig::cascade_asym}).   
Consider two sources $X$ and $Y$ that are jointly distributed according to $P_{X,Y}$.  
They will be described by separate encoders and passed to a single decoder in a cascade fashion.
Upon observing $X$, encoder $\mathrm{a}$ sends a message $M\in [\mathsf{L}_1]$ about $X$ to encoder $\mathrm{b}$. 
Encoder $\mathrm{b}$ then creates a final message $M' \in [\mathsf{L}_2]$ summarizing both sources $X$ and $Y$ and sends it to the decoder. 
We investigate the error probability $P_e$, which is the probability of the decoder recovers $\tilde{Z} \in \mathcal{Z}$ with the probability of excess distortion $P_e := \mathbf{P}\{d(X,Y,\tilde{Z}) > \mathsf{D} \}$, where $d: \mathcal{X} \times  \mathcal{Y}\times  \mathcal{Z}\rightarrow \mathbb{R}_{\geq 0}$ is a distortion measure. 
Due to the flexibility of the distortion function $d$, in general one can estimate any function of $X$ and $Y$ to tackle various objectives in practice~\cite{cuff2009cascade}. 

By Theorem~\ref{thm:det}, we bound $P_e$ by the following corollary. 

\mysmallskip

\begin{corollary}
\label{cor::cascade}
    Fix $P_{X,Y}$, $P_{U,V|X}$, function $z: \mathcal{U} \times \mathcal{V} \times \mathcal{Y} \rightarrow \mathcal{Z}$ and $\tilde{\mathsf{L}}_i, i=1,2,3$ with $\tilde{\mathsf{L}}_1 \tilde{\mathsf{L}}_2 \leq \mathsf{L}_1$ and $\tilde{\mathsf{L}}_2 \tilde{\mathsf{L}}_3 \leq \mathsf{L}_2$, there exists a deterministic coding scheme for the one-shot cascade multiterminal source coding problem such that the probability of excess distortion is bounded by
    \begin{align*}
        & P_e \\
        & \leq \mathbf{E}\bigg[\min\Big\{ 
        \mathbf{1}\{d(X, Y, Z)>\mathsf{D}\}  + 
        \gamma \tilde{\mathsf{L}}_1^{-1}\tilde{\mathsf{L}}_2^{-1} 2^{\iota(U,V; X | Y)} \\
        & \qquad \qquad \qquad + \gamma \tilde{\mathsf{L}}_1^{-1} 2^{-\iota(V; U,Y) + \iota(V; U,X)} \\
        & \qquad \qquad \qquad + 
        \tilde{\mathsf{L}}_2^{-1} 2^{- \iota(U; V, Y) + \iota(U; X)}
         \\
        &\qquad\qquad\qquad  + \gamma \tilde{\mathsf{L}}_3^{-1} 2^{\iota(Z;V,Y|U)} 
        \left(\tilde{\mathsf{L}}_2^{-1} 2^{\iota(U; X)} + 1 \right)  
        , \,\, 1
        \Big\}\bigg],
    \end{align*}
where $\gamma = \ln(|\mathcal{U}| \tilde{\mathsf{L}}_2 )+1$ and $X,Y,Z,U,V\sim P_{X}  P_{Y|X}  P_{U,V|X} P_{Z|Y,U,V}$. 
\end{corollary}

\mysmallskip

\begin{IEEEproof}
    We adapt the problem into our ADN framework by splitting the encoder $\mathrm{a}$, as shown in Figure~\ref{fig::cascade}. 
    The encoder $\mathrm{a1}$, encoder $\mathrm{a2}$, encoder $\mathrm{b}$ and decoder are referred to as nodes $1,2,3,4$, respectively.  
    
    \begin{figure*}[htpb]
    \centering \includegraphics[scale=0.42]{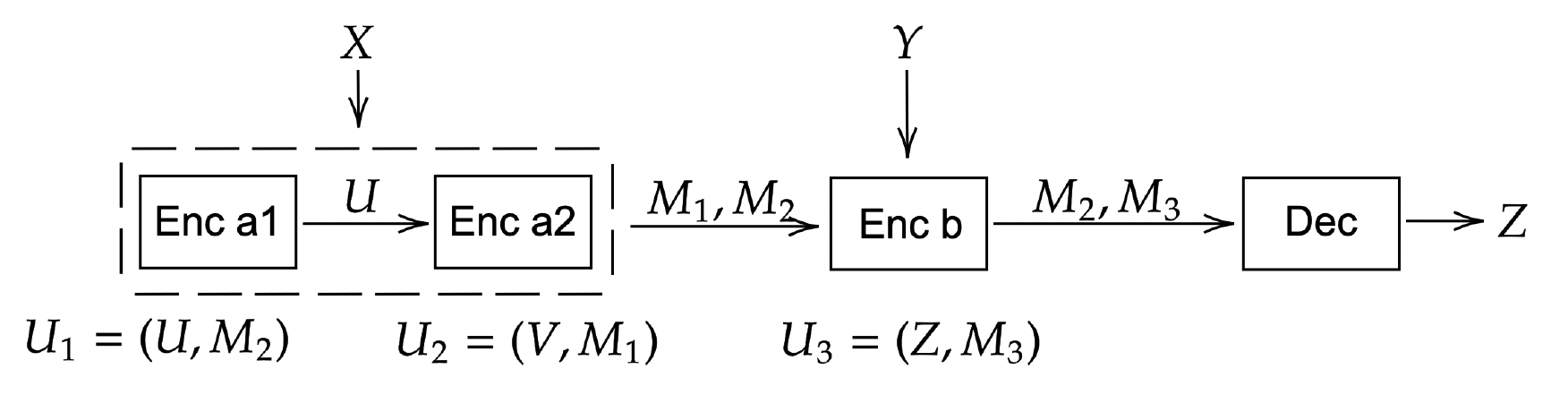}
    \caption{One-shot cascade multiterminal source coding in ADN framework by splitting the first encoder. } 
    \label{fig::cascade}
    \end{figure*}
    
    Let $M_i\in [\mathsf{L}_i]$ for $i=1,2,3$. 
    Encoder $\mathrm{a1}$ (node $1$) observes $X$, outputs $U$, and has an auxiliary $U_1 = (U, M_2)$. 
    Encoder $\mathrm{a2}$ (node $2$) observes $(U,X)$, outputs $(M_1, M_2)$, and has an auxiliary $U_2 = (V, M_1)$. 
    Encoder $\mathrm{b}$ (node $3$) observes $M_1$, $M_2$ and $Y$, outputs $(M_2, M_3)$, and has an auxiliary $U_3 = (Z, M_3)$. 
    The decoder observes $M_2, M_3$ and recovers $Z$ by using the function $z$ and our coding scheme.

    For each node $i=1,\ldots,4$ in the ADN, we describe its input $Y_i$, output $X_i$, auxiliary $U_i$ and the terms $B_{i,j}$ in Theorem~\ref{thm:det} as follows:
    \begin{enumerate}
        \item Node $1$ has input $Y_1 = X$, output $X_1 = U$ and auxiliary $U_1 = (U,M_2)$.
        
        \item Node $2$ has input $Y_2 = (U,X)$, output $X_2 = M_1$ and auxiliary $U_2 = (V,M_1)$. 
        
        \item Node $3$ has input $Y_3 = (Y, M_1, M_2)$, output $X_3 = M_3$, auxiliary $U_3 = (Z,M_3)$, and decodes with the order ``$U_2, U_1$'' (i.e., ``$U_{\mathrm{Enc}\, 1b}, U_{\mathrm{Enc}\, \mathrm{1a}}$''). 
        We have $d_3' = d_3 = 2$, and
        \begin{align*}
            B_{3, 1} & = \gamma \tilde{\mathsf{L}}_1^{-1}\tilde{\mathsf{L}}_2^{-1} 2^{\iota(U,V; X | Y)} + \\
            & \qquad \qquad \gamma \tilde{\mathsf{L}}_1^{-1} 2^{-\iota(V; U,Y) + \iota(V; U,X)}, \\
            B_{3, 2} & = \tilde{\mathsf{L}}_2^{-1} 2^{- \iota(U; V, Y) + \iota(U; X)},
        \end{align*} 
        where $\gamma = \ln(|\mathcal{U}| \tilde{\mathsf{L}}_2 )+1$. 
        
        \item Node $4$ has input $Y_4 = (M_2, M_3)$, output $X_4 = Z$, and decodes with the order ``$U_3, U_1 ?$'' (i.e., ``$U_{\mathrm{Enc}\, 2}, U_{\mathrm{Enc}\, \mathrm{1a}} ?$''). 
        By applying Theorem~\ref{thm:det} (note that $d_4' = 1$ and $d_4 = 2$), it gives 
        \begin{equation*}
            B_{4, 1} = \gamma \tilde{\mathsf{L}}_3^{-1} 2^{\iota(Z;V,Y|U)} 
            \left(\tilde{\mathsf{L}}_2^{-1} 2^{\iota(U; X)} + 1 \right),
        \end{equation*}
        where $\gamma = \ln(|\mathcal{U}| \tilde{\mathsf{L}}_2 )+1$. 
        
    \end{enumerate}

    Therefore, by applying Theorem~\ref{thm:det}, the probability of distortion exceeds the limit $P_e := \mathbf{P}\{d(X, Y, \tilde{Z}) > \mathsf{D} \}$ can be  bounded as the result stated in this corollary: 
    \begin{align*}
    & P_e \\
    & \leq \mathbf{E}\bigg[\min\Big\{ \mathbf{1}\{d(X, Y, Z)>\mathsf{D}\} + 
    B_{3,1} + B_{3,2} + B_{4,1}
    , 1\Big\}\bigg] \\
    & = \mathbf{E}\bigg[\min\Big\{ 
    \mathbf{1}\{d(X, Y, Z)>\mathsf{D}\}  + 
    \gamma \tilde{\mathsf{L}}_1^{-1}\tilde{\mathsf{L}}_2^{-1} 2^{\iota(U,V; X | Y)} \\
    & \qquad \qquad \qquad + \gamma \tilde{\mathsf{L}}_1^{-1} 2^{-\iota(V; U,Y) + \iota(V; U,X)} \\
    & \qquad \qquad \qquad + 
    \tilde{\mathsf{L}}_2^{-1} 2^{- \iota(U; V, Y) + \iota(U; X)}
     \\
    &\qquad\qquad\qquad  + \gamma \tilde{\mathsf{L}}_3^{-1} 2^{\iota(Z;V,Y|U)} 
    \left(\tilde{\mathsf{L}}_2^{-1} 2^{\iota(U; X)} + 1 \right)  
    , \,\, 1
    \Big\}\bigg],
    \end{align*}
    where $\gamma = \ln(|\mathcal{U}| \tilde{\mathsf{L}}_2 )+1$.

\end{IEEEproof}

Following the one-shot bound as shown above, we let $\tilde{\mathsf{L}}_i = 2^{n \tilde{R}_i}$ for $i=1,2,3$ and apply the law of large numbers. We obtain the asymptotic achievable region
\begin{align*}
    \tilde{R}_1 + \tilde{R}_2 & > I(X;U,V|Y), \\
    \tilde{R}_1               & > I(V; U, X) - I(V; U, Y), \\
    \tilde{R}_2               & > I(X;U) -  I(V, Y; U), \\
    \tilde{R}_2 + \tilde{R}_3 & > I(X;U) + I(Z;V,Y|U), \\
    \tilde{R}_3               & > I(Z;V,Y|U), 
\end{align*}
and $ D > \mathbf{E}[d(X,Y,Z)]$. 
By $\tilde{\mathsf{L}}_1 \tilde{\mathsf{L}}_2 \leq \mathsf{L}_1$ and $\tilde{\mathsf{L}}_2 \tilde{\mathsf{L}}_3 \leq \mathsf{L}_2$ and consider $R_1 = \tilde{R}_1 + \tilde{R}_2$ and $R_2 = \tilde{R}_2 + \tilde{R}_3$, by applying the Fourier-Motzkin elimination (using the PSITIP software~\cite{li2023automated}), we recover an asymptotic achievable region for the cascade multiterminal source coding problem: 
\begin{align*}
    R_1 & > I(X;U,V|Y), \\
    R_2 & > I(X;U)+ I(Z;V,Y|U),
\end{align*}
and $D > \mathbf{E}[d(X,Y,Z)]$, where $X,Y,Z,U,V\sim P_{X}  P_{Y|X}  P_{U,V|X} P_{Z|Y,U,V}$. 
This is the local-computing-and-forwarding inner bound in asymptotic case, as discussed in~\cite{cuff2009cascade} and also in~\cite[Section 21.4]{el2011network}.

\section{Examples of Acyclic Discrete Networks}
\label{sec::NIT}

In this section, to demonstrate the use of our main results, we apply  Theorem~\ref{thm::network_achievability} and Theorem~\ref{thm:det} on several settings in network information theory:  Gelfand-Pinsker~\cite{gelfand1980coding,Heegard1980}, Wyner-Ziv~\cite{wyner1976rate, wyner1978rate}, coding for computing~\cite{yamamoto1982wyner}, multiple access channels~\cite{ahlswede1971multi, liao1972multiple, ahlswede1974capacity} and broadcast channels~\cite{marton1979coding}, recovering similar results as the results in~\cite{li2021unified} and other works.

\subsection{Gelfand-Pinsker Problem}
\label{subsec::NIT_GP}
The one-shot version of the Gelfand-Pinsker problem~\cite{gelfand1980coding} is described as follows. 

\begin{figure}[htpb]
    \centering
    \includegraphics[scale=0.3]{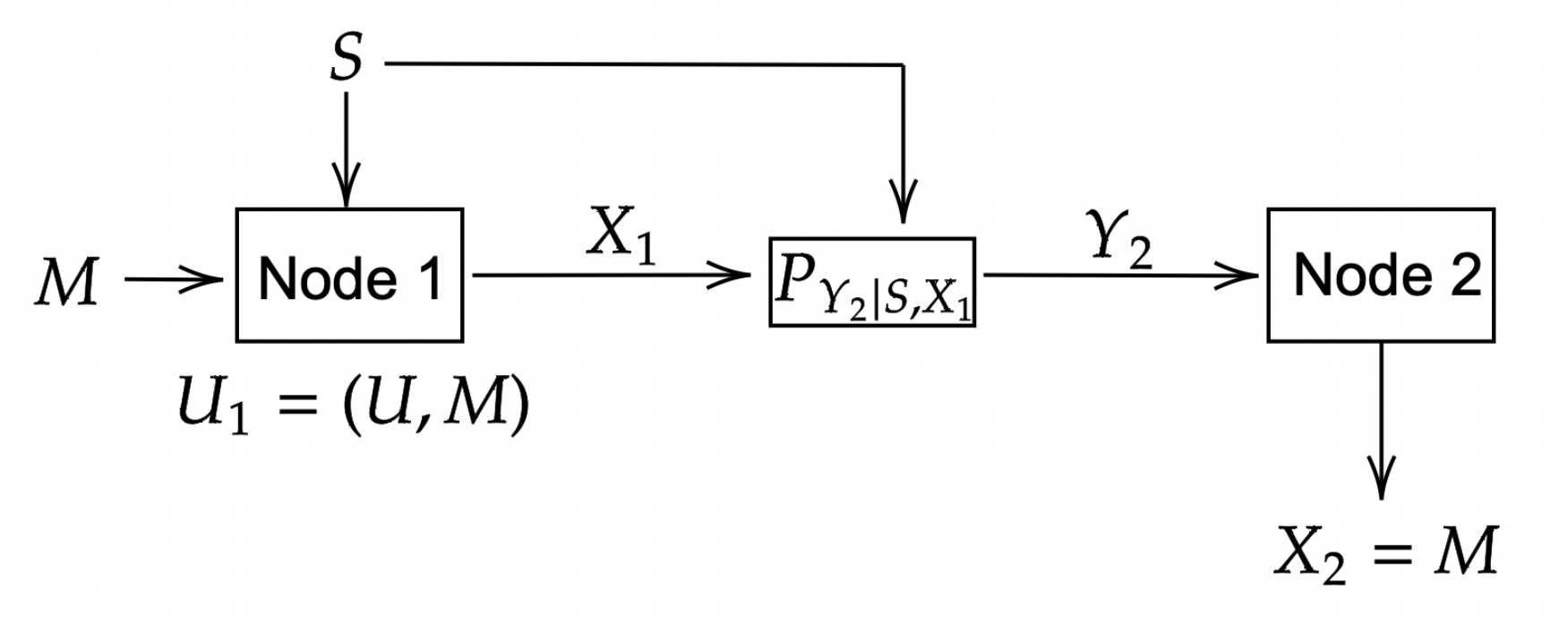}
    \caption{Gelfand-Pinsker problem in ADN framework.} 
    \label{fig::gp}
\end{figure}

Upon observing $M\sim \mathrm{Unif}[\mathsf{L}]$ and $S\sim P_S$, the encoder generates $X$ and sends $X$ through a channel $P_{Y|X,S}$. 
The decoder receives $Y$ and recovers $\hat{M}$. 
This can be considered as an ADN as follows (see Figure~\ref{fig::gp} for an illustration): 
in the ideal situation, let $Y_1 := (M,S)$ represent all the information coming into node $1$, $Y_2 := Y$, $P_{Y_2|Y_1, X_1}$ be $P_{Y|S, X}$, and $X_2 := M$.  
The auxiliary of node $1$ is $U_1 = (U,M)$ for some $U$ following $P_{U|S}$ given $S$. The decoding order of node $2$ is ``$U_1$'' (i.e., it only wants $U_1$). Since node $2$ has decoded $U_1$, $X_2$ is allowed to depend on $U_1=(U,M)$, and hence the choice $X_2 := M$ is valid in the ideal situation. Nevertheless, in the actual situation where we have $\tilde{X},\tilde{Y}$ instead of $X,Y$, the actual output $\tilde{X}_2$ will not be exactly $M$, though the error probability $P_e := \mathbf{P}(\tilde{X}_2 \neq M)$ can still be bounded. 
Applying Theorem~\ref{thm:det}, we obtain the following Corollary~\ref{cor::GP}.

\mysmallskip

\begin{corollary}
\label{cor::GP}
Fix $P_{U|S}$ and function $x:\mathcal{U}\times \mathcal{S}\rightarrow \mathcal{X}$. 
There exists a deterministic coding scheme for the channel $P_{Y|X,S}$ with state $S\sim P_S$ and message $M\sim \mathrm{Unif}[\mathsf{L}]$ such that
\begin{equation*}
	P_e \leq \mathbf{E}\big[  \min\big\{ 
	\mathsf{L}  2^{-\iota (U;Y) + \iota (U;S)}, 1 \big\}
	  \big], 
\end{equation*}
where $S,U,X,Y\sim P_S P_{U|S} \delta_{x(U,S)} P_{Y|X, S}$. 

\end{corollary}

\mysmallskip

This bound is similar to the one given in~\cite{li2021unified} (which is stronger than the one-shot bounds in~\cite{verdu2012non, yassaee2013technique, watanabe2015nonasymptotic} in the second order). 
They are the only known one-shot bounds that coincide with the best known second order result in~\cite{scarlett2015dispersions}, and are therefore stronger than the second order results implied by~\cite{verdu2012non, yassaee2013technique, watanabe2015nonasymptotic}.

\subsection{Wyner-Ziv Problem and Coding for Computing}
The Wyner-Ziv problem~\cite{wyner1976rate, wyner1978rate} in a one-shot setting is described as follows (see Figure~\ref{fig::wz} for an illustration).

Upon observing $X \sim P_X$, the encoder outputs $M \in [\mathsf{L}]$. 
The decoder receives $M$ and the side information $T \sim P_{T|X}$, and recovers $\tilde{Z}\in \mathcal{Z}$ with probability of excess distortion $P_e :=\mathbf{P} \{d(X,\tilde{Z})>\mathsf{D} \}$, where $d: \mathcal{X} \times\mathcal{Z} \rightarrow \mathbb{R}_{\geq 0} $ is a distortion measure. 
This can be considered as an ADN: in the ideal situation, $Y_1:= X$, $X_1 := M$, $Y_2 :=(M,T)$, $X_2 := Z$.
The auxiliary of node $1$ is $U_1 = (U,M)$ for some $U$ following $P_{U|X}$ given $X$. 
By Theorem~\ref{thm:det}, we bound  $P_e$ as follows.

\begin{figure}[htpb]
    \centering
    \includegraphics[scale=0.25]{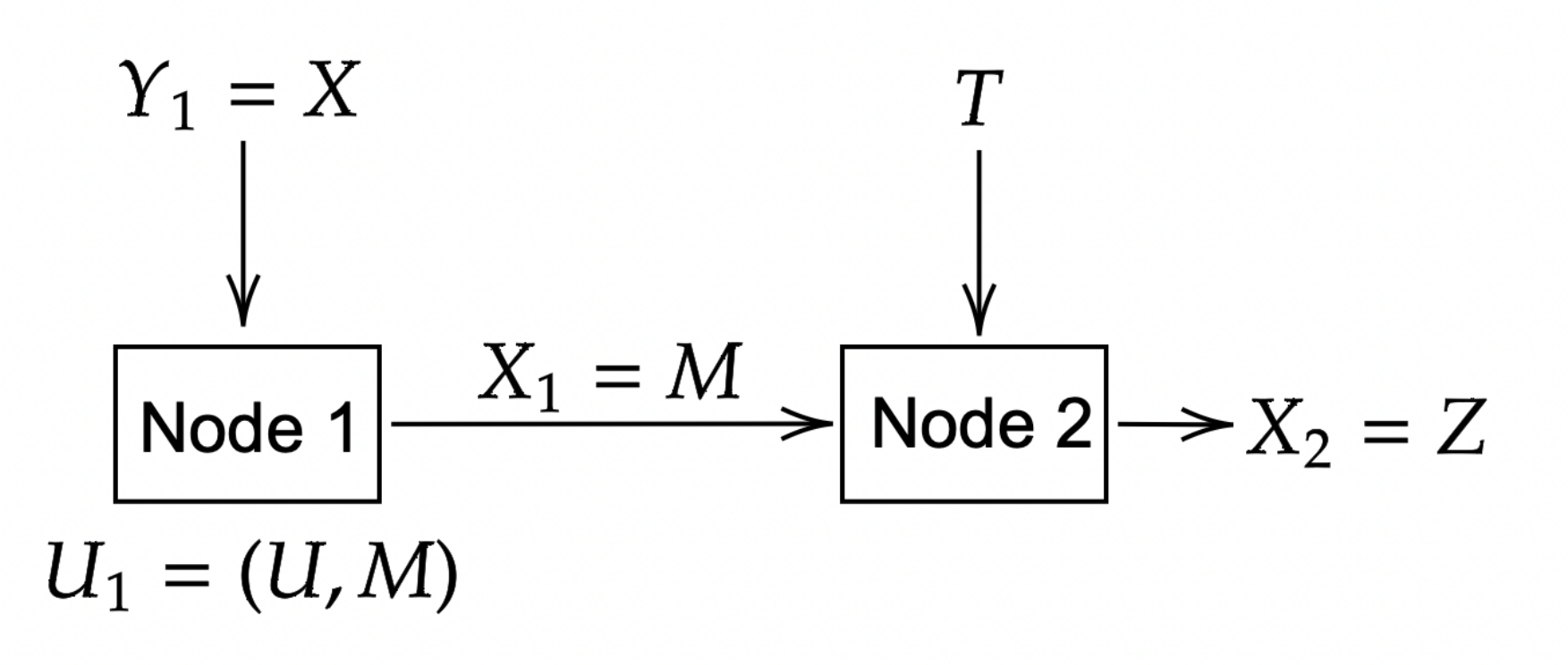}
    \caption{Wyner-Ziv problem in ADN framework.} 
    \label{fig::wz}
\end{figure}

\mysmallskip

\begin{corollary}
\label{cor::WZ}
Fix $P_{U|X}$ and function $z: \mathcal{U}\times \mathcal{Y}\rightarrow \mathcal{Z}$. 
There exists a deterministic coding scheme  for lossy source coding with source $X\sim P_X$, side information at the decoder $T\sim P_{T|X}$ and description $M\in [\mathsf{L}]$ such that 
\begin{equation}
	P_e \leq \mathbf{E}\Big[  \min\Big\{  
	\mathbf{1}\{d(X, Z)>\mathsf{D}\} + \mathsf{L}^{-1}  2^{-\iota (U;T) + \iota (U;X)}, 1 \Big\}
	  \Big], \label{eq:wyner_ziv}
\end{equation}
where $X,Y,U,Z \sim P_X P_{Y|X}  P_{U|X} \delta_{z(U,Y)}$. 
\end{corollary}

\mysmallskip

This bound is similar to the one given in~\cite{li2021unified}, which in turn is stronger than the one-shot bound in~\cite{verdu2012non}. 
This reduces to lossy source coding with $T=\emptyset$. 
Let $U=Z$, we have $P_e \leq \mathbf{P}(d(X, Z)>\mathsf{D}) + \mathbf{E}\left[ \min\left\{\mathsf{L}^{-1} 2^{\iota (Z;X)}, 1 \right\}\right]$.

We also consider coding for computing~\cite{yamamoto1982wyner}, where node $2$ recovers a function $f(X, T)$ of $X$ and $T$ with respect to distortion level $\mathsf{D}$ with a distortion measure $d(\cdot, \cdot)$. 
The probability of excess distortion is $P_e := \mathbf{P}\{d(f(X,T),\tilde{Z})>\mathsf{D} \}$.
We obtain a result similar to Corollary \ref{cor::WZ}, where \eqref{eq:wyner_ziv} is changed to \begin{align*}
    P_e 
    & \leq \mathbf{E}\big[  \min \big\{ \mathbf{1}\{d(f(X,T),Z)>\mathsf{D}\}\\
    & \qquad  \qquad \quad +  \mathsf{L}^{-1}  2^{-\iota (U;T) + \iota (U;X)}, \,\, 1 \big\}\big].
\end{align*}

\subsection{Multiple Access Channel}
The multiple access channel~\cite{ahlswede1971multi, liao1972multiple, ahlswede1974capacity} in a one-shot setting is described as follows.

There are two encoders, one decoder, and two independent messages $M_j\sim \mathrm{Unif}[\mathsf{L}_j]$ for $j=1,2$.
Encoder $j$ observes $M_j$ and creates $X_j$ for $j=1,2$. 
The decoder observes the output $Y$ of the channel $P_{Y |X_1, X_2}$ and produces the reconstructions $(\hat{M}_1, \hat{M}_2)$ of the messages. 
The error probability is defined as $P_e := \mathbf{P} \{(M_1, M_2)\neq (\hat{M}_1, \hat{M}_2) \}$. 
To consider this as an ADN, in the ideal situation, we let $Y_1 := M_1$, $Y_2 := M_2$, $Y_3 := Y$ and $X_3 := (M_1, M_2) $. 
We let $U_1 := (X_1, M_1)$ and $U_2 := (X_2, M_2)$. 
The decoding order of node $3$ is 
``$U_2, U_1$'' (i.e., decode $U_1$ (soft), and then $U_2$ (unique), and then $U_1$ (unique)). 
By Theorem~\ref{thm:det}, we have the following result. 

\begin{figure}[htpb]
    \centering
    \includegraphics[scale=0.3]{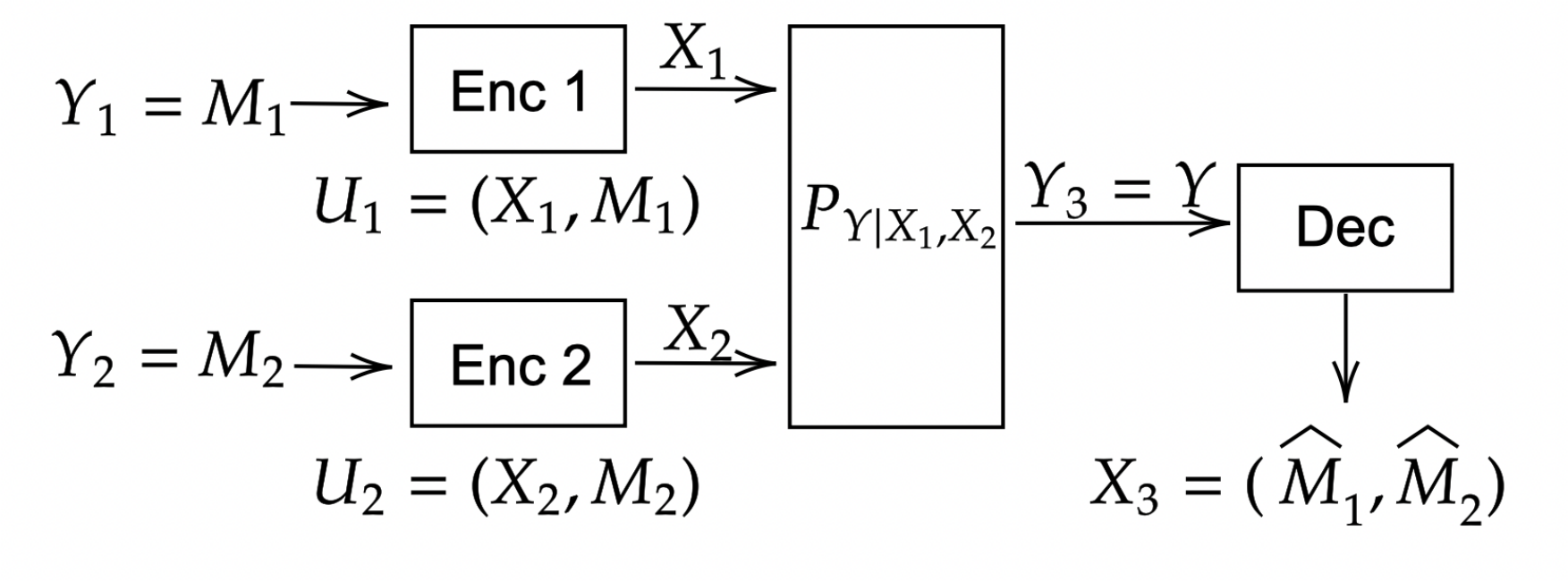}
    \caption{Multiple access channel in ADN framework.} 
    \label{fig::wz}
\end{figure}

\mysmallskip

\begin{corollary}
\label{cor::MAC_2}
    Fix $P_{X_1}, P_{X_2}$. 
    There exists a deterministic coding scheme for the multiple access channel $P_{Y |X_1, X_2}$ for messages $M_j\sim \mathrm{Unif}[1 : \mathrm{L}_j]$ for $j=1,2$ such that 
\begin{align*}
    P_e  
    & \leq  \mathbf{E}\Big[  \min \Big\{ \gamma \mathsf{L}_1 \mathsf{L}_2  2^{-\iota (X_1,X_2;Y)} + \\
    & \quad \quad \quad \gamma \mathsf{L}_2 2^{-\iota(X_2;Y|X_1)} +  \mathsf{L}_1
    2^{-\iota(X_1;Y|X_2)} ,1\Big\}
    \Big], 
\end{align*}
    where $\gamma := \ln (\mathsf{L}_1 |\mathcal{X}_1|)+1$, $(X_1, X_2, Y) \sim  P_{X_1} P_{X_2} P_{Y|X_1, X_2}$. 
\end{corollary}

\mysmallskip

This bound is similar to the one-shot bounds in~\cite{li2021unified,verdu2012non}. 
In the asymptotic setting, this will give the region $R_1 <I(X_1;Y|X_2) $, $ R_2 < I(X_2;Y|X_1)$, $ R_1 +R_2 < I(X_1,X_2;Y)$.

\subsection{Broadcast Channel with Private Messages}
The broadcast channel with private messages~\cite{marton1979coding} in a one-shot setting is described as follows.

Upon observing independent messages $M_j\sim \mathrm{Unif}[\mathsf{L}_j]$ for $j=1,2$, the encoder produces $X$ and sends it through a channel $P_{Y_1,Y_2|X}$. 
Decoder $j$ observes $Y_j$ and reconstructs $\hat{M}_j$ for $j=1,2$. 
By Theorem~\ref{thm:det}, we have the following result. 

\mysmallskip

\begin{corollary}
\label{cor::Broadcast_2}
Fix any $P_{U_1, U_2}$ and function $x: \mathcal{U}_1 \times \mathcal{U}_2 \rightarrow \mathcal{X}$. 
There exists a deterministic coding scheme for the broadcast channel $P_{Y_1, Y_2 |X}$  for independent messages $M_k \sim \mathrm{Unif}[\mathsf{L}_j]$ for $j = 1,2$, with the error probability bounded by 
\begin{align*}
    P_e  \leq \mathbf{E}\Big[\min\Big\{ \mathsf{L}_1  2^{-\iota (U_1; Y_1) } + \mathsf{L}_2  2^{-\iota (U_2;Y_2) + \iota (U_1; U_2)}, 1 \Big\} \Big],
\end{align*} 
where $(U_1,U_2, X, Y_1, Y_2)\sim P_{U_1, U_2} \delta _{x(U_1,U_2)} P_{Y_1, Y_2|X}$. 
\end{corollary} 

\mysmallskip

In the asymptotic case, this gives a corner point in Marton's region~\cite{marton1979coding}: $R_1 < I(U_1;Y_1), R_2 < I(U_2;Y_2)-I(U_1;U_2)$. 
Another corner point can be obtained by swapping the decoders. 

\section{Concluding Remarks}

In this paper, we have provided a unified one-shot coding framework over a general noisy network, applicable to any combination of source coding, channel
coding and coding for computing problems. 
A unified coding scheme may be useful for designing automated theorem proving tools for one-shot results. For example, \cite{li2023automated} gives an algorithm for deriving asymptotic inner and outer bounds for general ADMN~\cite{lee2018unified}. Extension of \cite{li2023automated} to one-shot results is left for future study. 
 
\section{Acknowledgements}

This work was partially supported by two grants from the Research Grants Council of the Hong Kong Special Administrative Region, China [Project No.s: CUHK 24205621 (ECS), CUHK 14209823 (GRF)].

The authors are grateful to Prof. Amin Gohari, Prof. Chandra Nair and Prof. Raymond W. Yeung for their valuable comments and suggestions at the early stage of this work. 
They would also like to thank the associate editor and the anonymous reviewers of this manuscript and the earlier conference version of this manuscript~\cite{liu2024one_isit} for their constructive and helpful comments.

\iflongver

\appendix

\subsection*{Proof of Theorem~\ref{thm::network_achievability} and Theorem~\ref{thm:det}\label{pf:thm}}

We first generate $N$ independent exponential processes $\mathbf{U}_i$ for $i\in [N]$ according to Section \ref{sec:expon}, which serve as the random codebooks. Each node $i$ will perform two steps: the \emph{decoding step} and the \emph{encoding step}. 

We describe the decoding step at node $i$. The node observes $Y_i$ and wants to decode $\overline{U}'_i = (U_{a_{i,j}})_{j \in [d'_i]}$, while utilizing $(U_{a_{i,j}})_{j \in [d'_i+1 .. d_i]}$ by non-unique decoding. 
For the sake of notational simplicity, we omit the subscript $i$ and write $d=d_i$, $d'=d'_i$, $a_k=a_{i,k}$, $\overline{U}_{\mathcal{S}}=\overline{U}_{i,\mathcal{S}} =(U_{a_{i,j}})_{j\in \mathcal{S}}$, $\overline{\mathbf{U}}_k := \mathbf{U}_{a_{i,k}}$. 
For each $j=1,\ldots,d'$, the node will perform soft decoding via the exponential process refinement (see Section \ref{sec:expon}) on $\overline{U}_{d}$, and then on $\overline{U}_{d-1}$, and so on up to $\overline{U}_{j+1}$, and then use all the distributions obtained to decode $\overline{U}_{j}$ uniquely using the Poisson functional representation. For example, when $d=3$, $d'=2$, the decoding process will be: $\overline{U}_{3}$ (soft), $\overline{U}_{2}$ (soft), $\overline{U}_{1}$ (unique), $\overline{U}_{3}$ (soft), $\overline{U}_{2}$ (unique). The choice of the sequence $a_{i,k}$ controls the decoding ordering of the random variables. The goal is to obtain the decoded variables $\hat{\overline{U}}_{1}, \ldots, \hat{\overline{U}}_{d'}$ that equal $\overline{U}_{1}, \ldots, \overline{U}_{d'}$ with high probability.

More precisely, for $j=1,\ldots,d'$, the node computes the decoded variable $\hat{\overline{U}}_j \in \overline{\mathcal{U}}_{j}$ by first computing the joint distributions $Q^{(j)}_{ \overline{U}_{[k..d]}}$ over  $\overline{\mathcal{U}}_{k}\times \ldots \times \overline{\mathcal{U}}_{d}$ for $k=d,d-1,\ldots,j+1$ recursively using the exponential process refinement as
\begin{align*}
 Q^{(j)}_{ \overline{U}_{[k..d]}} & :=  \\
 & \big( Q^{(j)}_{ \overline{U}_{[k+1..d]}} P_{\overline{U}_k| \overline{U}_{[k+1..d]}, \overline{U}_{[j-1]}, Y_i}(\cdot \,|\, \cdot ,\, \hat{\overline{U}}_{[j-1]}, Y_i)\big)^{\overline{\mathbf{U}}_k}, 
\end{align*}
i.e., first compute the semidirect product between $Q^{(j)}_{ \overline{U}_{[k+1..d]}}$ and the conditional distribution $P_{\overline{U}_k| \overline{U}_{[k+1..d]}, \overline{U}_{[j-1]}, Y_i}(\cdot \,|\, \cdot ,\, \hat{\overline{U}}_{[j-1]}, Y_i)$ (computed using the ideal joint distribution of $X^N,Y^N,U^N$) to obtain a distribution over $\overline{\mathcal{U}}_{k}\times \ldots \times \overline{\mathcal{U}}_{d}$, and then refine it by $\overline{\mathbf{U}}_k$ using Definition~\ref{def:refine}. For the base case, we assume $Q^{(j)}_{ \overline{U}_{[d+1..d]}}$ is the degenerate distribution. After we have computed $Q^{(j)}_{ \overline{U}_{[j+1..d]}}$, we can obtain $\hat{\overline{U}}_j$ using the Poisson functional representation~\eqref{eq:pfr} as $\hat{\overline{U}}_j= (\overline{\mathbf{U}}_j)_{\tilde{Q}^{(j)}_{\overline{U}_j}}$, where $\tilde{Q}^{(j)}_{\overline{U}_j}$ is the $\overline{U}_j$-marginal of
\begin{align}
&  Q^{(j)}_{ \overline{U}_{[j+1..d]}} P_{\overline{U}_j| \overline{U}_{[j+1..d]}, \overline{U}_{[j-1]}, Y_i}(\cdot \,|\, \cdot ,\, \hat{\overline{U}}_{[j-1]}, Y_i). \label{eq:Uj_pfr}
\end{align}
The node repeats this process for $j=1,\ldots,d'$ to obtain $\hat{\overline{U}}'_i=(\hat{\overline{U}}_{1}, \ldots, \hat{\overline{U}}_{d'})$. 

We then describe the encoding step at node $i$. It uses the Poisson functional representation (see Section \ref{sec:expon}) to obtain 
\begin{align}
& U_i = (\mathbf{U}_i)_{P_{U_i|Y_i, \overline{U}'_i}(\cdot | Y_i, \hat{\overline{U}}'_{i})}. \label{eq:Ui_pfr}
\end{align}
Finally, it generates $X_i$ from the conditional distribution $P_{X_i|Y_i,U_i, \overline{U}'_i}(\cdot | Y_i,U_i,\hat{\overline{U}}'_{i})$.

For the error analysis, we create a fictitious ``ideal network'' (with $N$ ``ideal nodes'') that is almost identical to the actual network. The only difference is that the ideal node $i$ uses the true $\overline{U}'_i$ (supplied by a genie) instead of the decoded $\hat{\overline{U}}'_i$ for the encoding step.
The random variables induced by the ideal network will have the same distribution as the ideal distribution of $X^N,Y^N,U^N$ in Theorem~\ref{thm::network_achievability}. Hence, we assume $X^N,Y^N,U^N$ are induced by the ideal network.
We couple the channels in the ideal network and the channels in the actual network, such that $Y_i=\tilde{Y}_i$ if $(X^{i-1},Y^{i-1})=(\tilde{X}^{i-1},\tilde{Y}^{i-1})$ (i.e., the ``channel noises'' in the two networks are the same).
If none of the actual nodes makes an error (i.e., $\hat{\overline{U}}'_i=\overline{U}'_i$ for all $i$), the actual network would coincide with the ideal network, and $(\tilde{X}^N,\tilde{Y}^N) = (X^N,Y^N)$. We consider the error probability conditional on $A:=(X^N,Y^N,U^N)$:
\[
F := \mathbf{P}\big(\exists \, i: \; \hat{\overline{U}}'_i=\overline{U}'_i\,\big|\,A \big).
\]
Note that $F$ is a random variable and is a function of $A=(X^N,Y^N,U^N)$. We have
\begin{align*}
F& =\mathbf{P}\big(\exists\, i\in [N], j \in [d'_i]:\, \hat{\overline{U}}_{i,j} \neq \overline{U}_{i,j} \,\big|\,A\big) \\
& = \sum_{i=1}^N \sum_{j=1}^{d'_i} \mathbf{P}\Big( \hat{\overline{U}}'_{[i-1]} = \overline{U}'_{[i-1]},\, \hat{\overline{U}}_{i,[j-1]} = \overline{U}_{i,[j-1]},\, \\
& \qquad\qquad\qquad \hat{\overline{U}}_{i,j} \neq \overline{U}_{i,j} \,\big|\,A\Big).
\end{align*}
For the term inside the summation (which is the probability that the first error we make is at $\overline{U}_{i,j}$), by \eqref{eq:Uj_pfr}, \eqref{eq:Ui_pfr} and the Poisson matching lemma~\cite{li2021unified} (we again omit the subscripts $i$ as in the description of the decoding step, e.g., we write $\overline{U}_{j} =\overline{U}_{i,j} =U_{a_{j}}=U_{a_{i,j}}$; we also simply write $P(\overline{U}_{j} | Y_{a_j}, \overline{U}'_{a_j})=P_{\overline{U}_{j}| Y_{a_j}, \overline{U}'_{a_j}}(\overline{U}_{j} | Y_{a_j}, \overline{U}'_{a_j})$), we have
\begin{align*}
 & \mathbf{P}\big(\hat{\overline{U}}'_{[i-1]}=\overline{U}'_{[i-1]},\,\hat{\overline{U}}_{i,[j-1]}=\overline{U}_{i,[j-1]},\,\hat{\overline{U}}_{i,j}\neq\overline{U}_{i,j} \,\big|\,A \big)\\
 & \le\mathbf{E}\bigg[\frac{P(\overline{U}_{j}|Y_{a_{j}},\overline{U}'_{a_{j}})}{Q^{(j)}(\overline{U}_{[j+1..d]})P(\overline{U}_{j}\,|\,\overline{U}_{[j+1..d]},\,\overline{U}_{[j-1]},Y_{i})} \,\bigg|\,A  \bigg]\\
 & \stackrel{(a)}{\le}\mathbf{E}\bigg[\frac{P(\overline{U}_{j}|Y_{a_{j}},\overline{U}'_{a_{j}})}{P(\overline{U}_{j}\,|\,\overline{U}_{[j+1..d]},\,\overline{U}_{[j-1]},Y_{i})} \\
 &\;\;\;\;\;\;\;\;\;\cdot\mathbf{E}\bigg[\frac{1}{Q^{(j)}(\overline{U}_{[j+1..d]})}\,\bigg|\,\overline{U}_{[d]},Y_{i},Y_{a_{j}},\overline{U}'_{a_{j}},\overline{\mathbf{U}}_{[j+1..d]}\bigg]\,\bigg|\,A\bigg]\\
 & \stackrel{(b)}{\le}\mathbf{E}\bigg[\frac{P(\overline{U}_{j}|Y_{a_{j}},\overline{U}'_{a_{j}})}{P(\overline{U}_{j}\,|\,\overline{U}_{[j+1..d]},\,\overline{U}_{[j-1]},Y_{i})}(\ln|\overline{\mathcal{U}}_{j+1}|+1)\\
 & \;\;\;\cdot\frac{1}{Q^{(j)}(\overline{U}_{[j+2..d]})}\bigg(\frac{P(\overline{U}_{j+1}|Y_{a_{j+1}},\overline{U}'_{a_{j+1}})}{P(\overline{U}_{j+1}|\overline{U}_{[j+2..d]},\overline{U}_{[j-1]},Y_{i})}+1\bigg)\,\bigg|\,A\bigg]\\
 & \stackrel{(c)}{\le}\mathbf{E}\bigg[\frac{P(\overline{U}_{j}|Y_{a_{j}},\overline{U}'_{a_{j}})}{P(\overline{U}_{j}\,|\,\overline{U}_{[j+1..d]},\,\overline{U}_{[j-1]},Y_{i})}\\
 &  \,\cdot\prod_{k=j+1}^{d'}(\ln|\overline{\mathcal{U}}_{k}|+1)\bigg(\frac{P(\overline{U}_{k}|Y_{a_{k}},\overline{U}'_{a_{k}})}{P(\overline{U}_{k}|\overline{U}_{[k+1..d]},\overline{U}_{[j-1]},Y_{i})}+1\bigg)\,\bigg|\,A\bigg]\\
&=B_{i,j}, 
\end{align*}
where (a) is by Jensen's inequality, (b) is due to Lemma \ref{lemma::PML2},
 (c) is by applying the same steps as (a) and (b) $d'-j-1$ times, and $\beta_{i,j}$ is given in \eqref{eq::beta}. 
 The proof of Theorem~\ref{thm::network_achievability} is completed by noting that $\delta_{\mathrm{TV}}(P_{X^N,Y^N},P_{\tilde{X}^N,\tilde{Y}^N}) \le \mathbf{P}((X^N,Y^N) \neq (\tilde{X}^N,\tilde{Y}^N)) \le \mathbf{E}[F]=\mathbf{E}[\min\{F,1\}]$. 

We now prove Theorem~\ref{thm:det}. Recall that the scheme we have constructed requires the public randomness $W$, which we have to fix in order to construct a deterministic coding scheme for Theorem~\ref{thm:det}. We have
\begin{align*}
&\mathbf{E}\big[\mathbf{P}\big((\tilde{X}^N,\tilde{Y}^N) \in \mathcal{E} \,\big|\, W \big)\big] \\
& = \mathbf{P}((\tilde{X}^N,\tilde{Y}^N) \in \mathcal{E}) \\
& \le \mathbf{P}\big((X^N,Y^N) \in \mathcal{E} \; \mathrm{or} \; (X^N,Y^N) \neq (\tilde{X}^N,\tilde{Y}^N)\big) \\
& = \mathbf{E}\big[ \mathbf{P}\big((X^N,Y^N) \in \mathcal{E} \; \mathrm{or} \; (X^N,Y^N) \neq (\tilde{X}^N,\tilde{Y}^N) \, \big|\, A\big) \big]\\
& \le \mathbf{E}\big[ \min\big\{ \mathbf{1}\{(X^N,Y^N) \in \mathcal{E}\}  \\
& \qquad \quad + \mathbf{P}\big((X^N,Y^N) \neq (\tilde{X}^N,\tilde{Y}^N) \, \big|\, A\big),\, 1 \big\} \big]\\
& \le \mathbf{E}\big[ \min\big\{ \mathbf{1}\{(X^N,Y^N) \in \mathcal{E}\} + F,\, 1 \big\} \big].
\end{align*}
Therefore, there exists a value $w$ such that $\mathbf{P}((\tilde{X}^N,\tilde{Y}^N) \in \mathcal{E} \,|\, W=w)$ satisfies the upper bound. Fixing the value of $W$ to $w$ gives a deterministic coding scheme.
\fi

\iflongver

\bibliographystyle{IEEEtran}
\bibliography{ref.bib}

\begin{thebibliography}{10}
\providecommand{\url}[1]{#1}
\csname url@samestyle\endcsname
\providecommand{\newblock}{\relax}
\providecommand{\bibinfo}[2]{#2}
\providecommand{\BIBentrySTDinterwordspacing}{\spaceskip=0pt\relax}
\providecommand{\BIBentryALTinterwordstretchfactor}{4}
\providecommand{\BIBentryALTinterwordspacing}{\spaceskip=\fontdimen2\font plus
\BIBentryALTinterwordstretchfactor\fontdimen3\font minus
  \fontdimen4\font\relax}
\providecommand{\BIBforeignlanguage}[2]{{%
\expandafter\ifx\csname l@#1\endcsname\relax
\typeout{** WARNING: IEEEtran.bst: No hyphenation pattern has been}%
\typeout{** loaded for the language `#1'. Using the pattern for}%
\typeout{** the default language instead.}%
\else
\language=\csname l@#1\endcsname
\fi
#2}}
\providecommand{\BIBdecl}{\relax}
\BIBdecl

\bibitem{el2011network}
A.~El~Gamal and Y.-H. Kim, \emph{Network information theory}.\hskip 1em plus
  0.5em minus 0.4em\relax Cambridge university press, 2011.

\bibitem{durisi2016toward}
G.~Durisi, T.~Koch, and P.~Popovski, ``Toward massive, ultrareliable, and
  low-latency wireless communication with short packets,'' \emph{Proceedings of
  the IEEE}, vol. 104, no.~9, pp. 1711--1726, 2016.

\bibitem{kostina2013lossy}
V.~Kostina and S.~Verd{\'u}, ``Lossy joint source-channel coding in the finite
  blocklength regime,'' \emph{IEEE Transactions on Information Theory},
  vol.~59, no.~5, pp. 2545--2575, 2013.

\bibitem{wang2011dispersion}
D.~{Wang}, A.~{Ingber}, and Y.~{Kochman}, ``The dispersion of joint
  source-channel coding,'' in \emph{2011 49th Annual Allerton Conference on
  Communication, Control, and Computing (Allerton)}, Sep. 2011, pp. 180--187.

\bibitem{tan2013dispersions}
V.~Y. Tan and O.~Kosut, ``On the dispersions of three network information
  theory problems,'' \emph{IEEE Transactions on Information Theory}, vol.~60,
  no.~2, pp. 881--903, 2013.

\bibitem{polyanskiy2010channel}
Y.~Polyanskiy, H.~V. Poor, and S.~Verd{\'u}, ``Channel coding rate in the
  finite blocklength regime,'' \emph{IEEE Trans. Inf. Theory}, vol.~56, no.~5,
  pp. 2307--2359, 2010.

\bibitem{liu2024oblivious_isit}
Y.~Liu, S.~H. Advary, and C.~T. Li, ``Nonasymptotic oblivious relaying and
  variable-length noisy lossy source coding,'' in \emph{2025 IEEE International
  Symposium on Information Theory (ISIT)}.\hskip 1em plus 0.5em minus
  0.4em\relax IEEE, 2025.

\bibitem{feinstein1954new}
A.~Feinstein, ``A new basic theorem of information theory,'' \emph{IRE Trans.
  Inf. Theory}, no.~4, pp. 2--22, 1954.

\bibitem{shannon1957certain}
C.~E. Shannon, ``Certain results in coding theory for noisy channels,''
  \emph{Information and control}, vol.~1, no.~1, pp. 6--25, 1957.

\bibitem{hayashi2009information}
M.~Hayashi, ``Information spectrum approach to second-order coding rate in
  channel coding,'' \emph{IEEE Transactions on Information Theory}, vol.~55,
  no.~11, pp. 4947--4966, 2009.

\bibitem{verdu2012non}
S.~Verd{\'u}, ``Non-asymptotic achievability bounds in multiuser information
  theory,'' in \emph{2012 50th Annual Allerton Conference on Communication,
  Control, and Computing (Allerton)}.\hskip 1em plus 0.5em minus 0.4em\relax
  IEEE, 2012, pp. 1--8.

\bibitem{yassaee2013technique}
M.~H. Yassaee, M.~R. Aref, and A.~Gohari, ``A technique for deriving one-shot
  achievability results in network information theory,'' in \emph{2013 IEEE
  International Symposium on Information Theory}.\hskip 1em plus 0.5em minus
  0.4em\relax IEEE, 2013, pp. 1287--1291.

\bibitem{liu2015one}
J.~Liu, P.~Cuff, and S.~Verd{\'u}, ``One-shot mutual covering lemma and
  {M}arton's inner bound with a common message,'' in \emph{2015 IEEE
  International Symposium on Information Theory (ISIT)}.\hskip 1em plus 0.5em
  minus 0.4em\relax IEEE, 2015, pp. 1457--1461.

\bibitem{song2016likelihood}
E.~C. Song, P.~Cuff, and H.~V. Poor, ``The likelihood encoder for lossy
  compression,'' \emph{IEEE Transactions on Information Theory}, vol.~62,
  no.~4, pp. 1836--1849, 2016.

\bibitem{watanabe2015nonasymptotic}
S.~Watanabe, S.~Kuzuoka, and V.~Y. Tan, ``Nonasymptotic and second-order
  achievability bounds for coding with side-information,'' \emph{IEEE
  Transactions on Information Theory}, vol.~61, no.~4, pp. 1574--1605, 2015.

\bibitem{yassaee2013non}
M.~H. Yassaee, M.~R. Aref, and A.~Gohari, ``Non-asymptotic output statistics of
  random binning and its applications,'' in \emph{2013 IEEE International
  Symposium on Information Theory}.\hskip 1em plus 0.5em minus 0.4em\relax
  IEEE, 2013, pp. 1849--1853.

\bibitem{li2021unified}
C.~T. Li and V.~Anantharam, ``A unified framework for one-shot achievability
  via the poisson matching lemma,'' \emph{IEEE Transactions on Information
  Theory}, vol.~67, no.~5, pp. 2624--2651, 2021.

\bibitem{verdu1994general}
S.~Verd{\'u} and T.~S. Han, ``A general formula for channel capacity,''
  \emph{IEEE Trans. Inf. Theory}, vol.~40, no.~4, pp. 1147--1157, 1994.

\bibitem{liu2024hiding}
Y.~Liu and C.~T. Li, ``One-shot information hiding,'' in \emph{accepted at the
  IEEE Information Theory Workshop}.\hskip 1em plus 0.5em minus 0.4em\relax
  IEEE, 2024.

\bibitem{khisti2024unequal}
A.~Khisti, A.~Behboodi, G.~Cesa, and P.~Kumar, ``Unequal message protection:
  One-shot analysis via poisson matching lemma,'' in \emph{2024 IEEE
  International Symposium on Information Theory (ISIT)}.\hskip 1em plus 0.5em
  minus 0.4em\relax IEEE, 2024.

\bibitem{guo2024hypothesis}
Y.~Guo, S.~Salehkalaibar, S.~C. Draper, and W.~Yu, ``One-shot achievability
  region for hypothesis testing with communication constraint,'' in
  \emph{accepted at the IEEE Information Theory Workshop}.\hskip 1em plus 0.5em
  minus 0.4em\relax IEEE, 2024.

\bibitem{hentila2024communication}
H.~Hentil{\"a}, Y.~Y. Shkel, and V.~Koivunen, ``Communication-constrained
  secret key generation: Second-order bounds,'' \emph{IEEE Transactions on
  Information Theory}, 2024.

\bibitem{li2018strong}
C.~T. Li and A.~El~Gamal, ``Strong functional representation lemma and
  applications to coding theorems,'' \emph{IEEE Transactions on Information
  Theory}, vol.~64, no.~11, pp. 6967--6978, 2018.

\bibitem{lei2022neural}
E.~Lei, H.~Hassani, and S.~S. Bidokhti, ``Neural estimation of the
  rate-distortion function with applications to operational source coding,''
  \emph{IEEE Journal on Selected Areas in Information Theory}, vol.~3, no.~4,
  pp. 674--686, 2022.

\bibitem{li2020minimax}
C.~T. Li, X.~Wu, A.~Ozgur, and A.~El~Gamal, ``Minimax learning for distributed
  inference,'' \emph{IEEE Transactions on Information Theory}, vol.~66, no.~12,
  pp. 7929--7938, 2020.

\bibitem{liu2024universal}
Y.~Liu, W.-N. Chen, A.~{\"O}zg{\"u}r, and C.~T. Li, ``Universal exact
  compression of differentially private mechanisms,'' \emph{Advances in Neural
  Information Processing Systems}, 2024.

\bibitem{liu2024one_isit}
Y.~Liu and C.~T. Li, ``One-shot coding over general noisy networks,'' in
  \emph{2024 IEEE International Symposium on Information Theory (ISIT)}, 2024,
  pp. 3124--3129.

\bibitem{lee2018unified}
S.-H. Lee and S.-Y. Chung, ``A unified random coding bound,'' \emph{IEEE
  Transactions on Information Theory}, vol.~64, no.~10, pp. 6779--6802, 2018.

\bibitem{kim2007coding}
Y.-H. Kim, ``Coding techniques for primitive relay channels,'' in \emph{Proc.
  Forty-Fifth Annual Allerton Conf. Commun., Contr. Comput}, 2007, p. 2007.

\bibitem{mondelli2019new}
M.~Mondelli, S.~H. Hassani, and R.~Urbanke, ``A new coding paradigm for the
  primitive relay channel,'' \emph{Algorithms}, vol.~12, no.~10, p. 218, 2019.

\bibitem{el2021achievable}
A.~El~Gamal, A.~Gohari, and C.~Nair, ``Achievable rates for the relay channel
  with orthogonal receiver components,'' in \emph{2021 IEEE Information Theory
  Workshop (ITW)}.\hskip 1em plus 0.5em minus 0.4em\relax IEEE, 2021, pp. 1--6.

\bibitem{el2022strengthened}
------, ``A strengthened cutset upper bound on the capacity of the relay
  channel and applications,'' \emph{IEEE Transactions on Information Theory},
  vol.~68, no.~8, pp. 5013--5043, 2022.

\bibitem{gelfand1980coding}
S.~I. Gel'fand and M.~S. Pinsker, ``Coding for channel with random
  parameters,'' \emph{Probl. Contr. and Inf. Theory}, vol.~9, no.~1, pp.
  19--31, 1980.

\bibitem{Heegard1980}
C.~Heegard and A.~El~Gamal, ``On the capacity of computer memory with
  defects,'' \emph{IEEE Transactions on Information Theory}, vol.~29, no.~5,
  pp. 731--739, 1983.

\bibitem{el2005relay}
A.~El~Gamal and N.~Hassanpour, ``Relay-without-delay,'' in \emph{Proceedings.
  International Symposium on Information Theory, 2005. ISIT 2005.}\hskip 1em
  plus 0.5em minus 0.4em\relax IEEE, 2005, pp. 1078--1080.

\bibitem{el2007relay}
A.~El~Gamal, N.~Hassanpour, and J.~Mammen, ``Relay networks with delays,''
  \emph{IEEE Transactions on Information Theory}, vol.~53, no.~10, pp.
  3413--3431, 2007.

\bibitem{wyner1976rate}
A.~Wyner and J.~Ziv, ``The rate-distortion function for source coding with side
  information at the decoder,'' \emph{IEEE Transactions on information Theory},
  vol.~22, no.~1, pp. 1--10, 1976.

\bibitem{wyner1978rate}
A.~D. Wyner, ``The rate-distortion function for source coding with side
  information at the decoder-ii. general sources,'' \emph{Information and
  control}, vol.~38, no.~1, pp. 60--80, 1978.

\bibitem{yamamoto1982wyner}
H.~Yamamoto, ``Wyner-ziv theory for a general function of the correlated
  sources (corresp.),'' \emph{IEEE Transactions on Information Theory},
  vol.~28, no.~5, pp. 803--807, 1982.

\bibitem{ahlswede1971multi}
R.~Ahlswede, ``Multi-way communication channels,'' in \emph{2nd Int. Symp.
  Inform. Theory, Tsahkadsor, Armenian SSR}, 1971, pp. 23--52.

\bibitem{liao1972multiple}
H.~Liao, ``Multiple access channels,'' Ph.D. dissertation, University of
  Hawaii, Honolulu, HI, 1972.

\bibitem{ahlswede1974capacity}
R.~Ahlswede, ``The capacity region of a channel with two senders and two
  receivers,'' \emph{The annals of probability}, vol.~2, no.~5, pp. 805--814,
  1974.

\bibitem{marton1979coding}
K.~Marton, ``A coding theorem for the discrete memoryless broadcast channel,''
  \emph{IEEE Transactions on Information Theory}, vol.~25, no.~3, pp. 306--311,
  1979.

\bibitem{cuff2009cascade}
P.~Cuff, H.-I. Su, and A.~El~Gamal, ``Cascade multiterminal source coding,'' in
  \emph{2009 IEEE International Symposium on Information Theory}.\hskip 1em
  plus 0.5em minus 0.4em\relax IEEE, 2009, pp. 1199--1203.

\bibitem{shannon1961two}
C.~E. Shannon, ``Two-way communication channels,'' in \emph{Proceedings of the
  Fourth Berkeley Symposium on Mathematical Statistics and Probability, Volume
  1: Contributions to the Theory of Statistics}, vol.~4.\hskip 1em plus 0.5em
  minus 0.4em\relax University of California Press, 1961, pp. 611--645.

\bibitem{lim2011noisy}
S.~H. Lim, Y.-H. Kim, A.~El~Gamal, and S.-Y. Chung, ``Noisy network coding,''
  \emph{IEEE Transactions on Information Theory}, vol.~57, no.~5, pp.
  3132--3152, 2011.

\bibitem{van1971three}
E.~C. Van Der~Meulen, ``Three-terminal communication channels,'' \emph{Advances
  in applied Probability}, vol.~3, no.~1, pp. 120--154, 1971.

\bibitem{cover1979capacity}
T.~Cover and A.~E. Gamal, ``Capacity theorems for the relay channel,''
  \emph{IEEE Transactions on information theory}, vol.~25, no.~5, pp. 572--584,
  1979.

\bibitem{gumbel1954statistical}
E.~J. Gumbel, ``Statistical theory of extreme valuse and some practical
  applications,'' \emph{Nat. Bur. Standards Appl. Math. Ser. 33}, 1954.

\bibitem{huijben2022review}
I.~A. Huijben, W.~Kool, M.~B. Paulus, and R.~J. Van~Sloun, ``A review of the
  gumbel-max trick and its extensions for discrete stochasticity in machine
  learning,'' \emph{IEEE Transactions on Pattern Analysis and Machine
  Intelligence}, vol.~45, no.~2, pp. 1353--1371, 2022.

\bibitem{nair2009capacity}
C.~Nair and A.~El~Gamal, ``The capacity region of a class of three-receiver
  broadcast channels with degraded message sets,'' \emph{IEEE Transactions on
  Information Theory}, vol.~55, no.~10, pp. 4479--4493, 2009.

\bibitem{chong2008han}
H.-F. Chong, M.~Motani, H.~K. Garg, and H.~El~Gamal, ``On the {H}an-{K}obayashi
  region for the interference channel,'' \emph{IEEE Transactions on Information
  Theory}, vol.~54, no.~7, pp. 3188--3195, 2008.

\bibitem{bennett2002entanglement}
C.~H. Bennett, P.~W. Shor, J.~Smolin, and A.~V. Thapliyal,
  ``Entanglement-assisted capacity of a quantum channel and the reverse
  {S}hannon theorem,'' \emph{IEEE Trans. Inf. Theory}, vol.~48, no.~10, pp.
  2637--2655, 2002.

\bibitem{cuff2013distributed}
P.~Cuff, ``Distributed channel synthesis,'' \emph{IEEE Transactions on
  Information Theory}, vol.~59, no.~11, pp. 7071--7096, 2013.

\bibitem{cuff2010coordination}
P.~Cuff, H.~Permuter, and T.~M. Cover, ``Coordination capacity,'' \emph{IEEE
  Trans. Inf. Theory}, vol.~56, no.~9, pp. 4181--4206, Sept 2010.

\bibitem{li2023automated}
C.~T. Li, ``An automated theorem proving framework for information-theoretic
  results,'' \emph{IEEE Transactions on Information Theory}, vol.~69, no.~11,
  pp. 6857--6877, 2023.

\bibitem{berger1978multiterminal}
T.~Berger, ``Multiterminal source coding,'' in \emph{The Information Theory
  Approach to Communications}, G.~Longo, Ed.\hskip 1em plus 0.5em minus
  0.4em\relax New York: Springer-Verlag, 1978, pp. 171--231.

\bibitem{tung1978multiterminal}
S.-Y. Tung, ``Multiterminal source coding,'' Ph.D. dissertation, Cornell
  University, Ithaca, NY, 1978.

\bibitem{scarlett2015dispersions}
J.~Scarlett, ``On the dispersions of the {G}el’fand--{P}insker channel and
  dirty paper coding,'' \emph{IEEE Transactions on Information Theory},
  vol.~61, no.~9, pp. 4569--4586, 2015.

\end{thebibliography}
\fi

\end{document}